\documentclass[12pt,a4paper]{article}
\usepackage{amsmath,amsthm,amsfonts,amssymb,amscd}
\usepackage{wasysym}
\usepackage{graphicx}
\usepackage{wrapfig}

\newtheorem{theorem}{Theorem}[section]

\newtheorem{proposition}{Proposition}[section]
\newtheorem{lemma}{Lemma}[section]

\newcommand{\Irm}{ {\,\rm I}}
\newcommand{\erm}{{\rm e}}

\newcommand{\ov} {\overline}

\newcommand{\Io}{{\rm I}_\circ}

\newcommand{\Oc} {{\cal O}}

\newcommand{\ap} {\alpha}
\newcommand{\bt} {\beta}
\newcommand{\dt} {\delta}

\newcommand{\ep} {\epsilon}
\newcommand{\Gm} {\Gamma}
\newcommand{\gm} {\gamma}
\newcommand{\Lb} {\Lambda}
\newcommand{\lb} {\lambda}
\newcommand{\om} {\omega}
\newcommand{\Om} {\Omega}
\newcommand{\sg} {\sigma}

\newcommand{\defi}{\stackrel{\rm def}{=}}

\def\R{{\mathbb R}}

\newcommand{\I}{\mathbf {I}  }
\newcommand{\bproof}{\noindent {\it Proof. }}
\newcommand{\eproof}{\hfill$\Box$}
\newcommand{\nd} {\noindent}

\begin{document}
\begin{titlepage}

  \hrule
\smallskip
\nd  
Published in: S\~ao Paulo Journal of Mathematical Sciences (2020) v.  14,
p. 1-48 \\
\nd https://doi.org/10.1007/s40863-019-00162-3
\smallskip
\hrule

\begin{center}

  {\LARGE  The  theory of figures of Clairaut
    with focus on the gravitational  modulus: inequalities and
  an improvement in the Darwin-Radau equation.}

\end{center}
 
\vskip  1.0truecm
\centerline {{\large Clodoaldo Ragazzo}}

\vskip 0.5truecm

\centerline { {\sl  Instituto de Matem\'atica e Estat\'\i stica}}
\centerline {{\sl Universidade de S\~ao Paulo}}

\vskip .5truecm

\begin{abstract} 
This paper contains a  review of  Clairaut's theory with focus on the determination
of a gravitational  modulus $\gm$ defined as 
$\left(\frac{C-\Io}{\Io}\right)\gm=\frac{2}{3}\Om^2$, where $C$ and $\Io$
are the polar and mean moment of inertia of the body and $\Om$ is the body spin.
  The constant $\gm$ is  related to the static fluid Love number 
  $k_2= \frac{3\Io G}{R^5} \frac{1}{\gm}$, where $R$ is the body radius and $G$ is the
  gravitational constant. 
  The new results are: a  variational principle
  for $\gm$,  upper and lower bounds on the ellipticity that improve
  previous bounds by Chandrasekhar  \cite{chandra1963},
  and a semi-empirical procedure for estimating $\gm$
  from the knowledge of $m$, $\Io$, and $R$,
  where $m$ is the mass of the body. The main conclusion is that
  for $0.2\le \Io/(mR^2)\le 0.4$ the approximation 
  $\gm\approx G \sqrt{ \frac{2^7}{5^5}\frac{m^5}{\Io^3}}\defi \gm_\Irm$
  is a better estimate for $\gm$
  than that obtained from the
  Darwin-Radau equation, denoted as $\gm_{DR}$.
  Moreover,  an inequality in the paper implies that the 
  Darwin-Radau approximation may  be valid only  for  $0.3\le \Io/(mR^2)\le 0.4$ and
  within this range 
   $|\gm_{DR}/\gm_\Irm -1|<0.14\% $. 
\end{abstract} 

\vskip .5truecm

\noindent



\vfill
\hrule
\smallskip
\nd 
CR is partially supported by FAPESP 2016/25053-8.

\nd  e-mail:ragazzo@usp.br
\end{titlepage}

\pagebreak
\section{Introduction}
\label{intr}

Many celestial bodies are large extended objects,  almost spherical in shape, 
almost rigid, and usually with spin.
In most cases the deformation is caused by the motion itself
and it is difficult to determine  the laws that relate motion to deformation.
This paper is about one of these laws: the static gravitational rigidity
that relates the spin of a body to its  flatness.
The static gravitational
rigidity is characterized by a single parameter \cite{rr2015} \cite{rr2017}:
{\it the gravitational modulus} $\gm$, which is closely related to 
the static fluid Love number $k_2$.
The goal of this paper is to review the classical theory used in the determination
of this parameter. The novelties are: the unusual
approach to the subject in considering the  mean moment of inertia
of the body 
as a given quantity (usually this is the main  unknown to be determined),
a new variational principle for the static gravitational rigidity,  and a new
differential equation for $\gm$ that allows
for obtaining sharp upper and lower bounds for this quantity.
These three new aspects are detailed in the following paragraphs.

A celestial body is considered large  when self-gravitation dominates
over any  elastic stress. In this case the self-gravitational
force is  balanced by the hydrostatic pressure  and at rest  the resulting   
equilibrium shape  is spherically symmetric with moment of inertia
$\Io$ along any axis passing through the center of mass.
We are interested in small  deformations of this spherical  shape under
 slow  rotational motion. 
 {\it Deformations are always assumed to be  small and incompressible.}\footnote{For
   small 
  deformations the incompressibility hypothesis is quite
  reasonable because  the modulus
  of compressibility of solids and fluids is much larger than the modulus of shear.}
Under these conditions  it was showed by G. Darwin \cite{Rochester-Smylie} that  the
trace of the moment
of inertia tensor remains invariant, namely, if
$A\le B\le C$ denote the time-dependent principal moments of inertia of
 a body then 
\begin{equation}
  \Io=\frac{A+B+C}{3} \label{Iodef}
\end{equation}
does not depend on time.

To a given body of  mass $m$  and  moment of inertia $\Io$ we can associate 
the {\it radius of inertia} that is  a length scale  given by
\begin{equation}
R_I=\sqrt{5\Io/(2m)}
\label{RI}
\end{equation}
The radius of inertia is the radius of a homogeneous
ball of mass $m$ and moment of inertia $\Io$.
The constants $m$, $\Io$ and $G$ (the gravitational constant) also define a time
scale by means of
\begin{equation}
\gm_\Irm=\om_\Irm^2=\frac{4}{5}\frac{Gm}{R_\Irm^3}=
  \frac{2\Io G}{R_\Irm^5}
\label{gmI}
\end{equation}
The quantity $\om_{\Irm}$, which we call {\it the inertial frequency},
admits three interpretations:
it is the smallest angular  frequency of oscillations
of a homogeneous  
spherical mass of liquid  with  mass $m$ and  moment of inertia $\Io$
(\cite{Lamb} paragraph 262 Eq. (10)),  it is $4/5$ times the square of 
the angular velocity of a particle moving with a circular orbit of radius
$R_\Irm$ around a point mass $m$, and it is the gravitational modulus,
to be defined below, of a homogeneous  
spherical mass of liquid  with  mass $m$ and radius $R_\Irm$.
The three quantities $m$, $R_\Irm$,  and $\gm_\Irm$  do not depend on the
deformations of the body  and may be considered as invariant
properties of the given deformable body.

This entire paper is about the determination of the gravitational modulus
 $\gm$ that is defined in the
following way. The effect of the centrifugal force upon an isolated  body under
uniform steady rotation
 is to  flatten the body  along the axis of
 rotation. Let $\Om>0$ be the constant angular spin rate.\footnote{A main issue in the
   dynamics of deformable bodies is to define a ``body frame'' or, equivalently,
   a notion of body rotation. In principle, each point of the body may rotate
   differently about a given inertial frame. The definition adopted in \cite{rr2015}
   and \cite{rr2017} is that of Tisserand: if $L$ is the instantaneous
   angular momentum (vector) and $\I$ is the instantaneous moment of inertia
   (matrix) then the instantaneous angular velocity (vector) $\Om$ is defined by
   $L=\I \Om$. The Sun, for instance, requires the use of this definition.}
 The amount of flatness can be measured by the increase in the
moment of inertia $C$ along the axis of rotation.
In \cite{rr2015} and \cite{rr2017}, {\it under the hypothesis that}
\begin{equation}
\frac{C-\Io}{\Io}\quad
\text{\it is small},
\label{smallstrain}
\end{equation}
the gravitational modulus 
$\gm$ relating the rotation rate to  the body flatness was defined as
\begin{equation}
\left(\frac{C-\Io}{\Io}\right)\gm=\frac{2}{3}\Om^2
\label{gmdef}
\end{equation}
(in the notation of \cite{rr2015} $\frac{C-\Io}{\Io}=-B_{33}$).
The factor $\frac{C-\Io}{\Io}$ in equation (\ref{gmdef}) represents the
moment of inertia strain (nondimensional) while $\frac{2}{3}\Om^2$ (time$^{-2}$)
represents the centrifugal stress, therefore the gravitational modulus $\gm$
has the unusual dimension time$^{-2}$.
Notice that the definition of $\gm$ makes sense even when $\frac{C-\Io}{\Io}$ and 
$\Om^2$ are large but, in this case,   $\gm$ may be a nonlinear function of $\Om^2$
\cite{zharkov}. So, in order to keep the relation between $\frac{C-\Io}{\Io}$ and  
$\Om^2$ within the linear range we impose
the smallness of $\frac{C-\Io}{\Io}$, namely hypothesis (\ref{smallstrain}).
Equivalently, 
we may impose the smallness of the nondimensional quantity
\begin{equation}    \zeta_\Irm=\frac{\Om^2R_\Irm^3}{Gm}\label{muzeta}
  \end{equation}
  which is the ratio of the centrifugal to the gravitational  acceleration  at
  $R_\Irm$.
  The gravitational modulus $\gm$,
  which is constant, fails to satisfy equation (\ref{gmdef})
  as $\zeta_\Irm$ becomes large.

Notice that the geometric radius of the body $R$ ({\it in this paper
  $R$ always refer to the volumetric mean radius}) does not appear in the definition
of $\gm$.\footnote{
  Our parameter $\gm$ is closely related to the inverse of the parameter $\Lb_2$
  in equation (37.6) of \cite{zharkov}, the difference is that their definition uses
  the radius of the body.}
 
Using
\[  J_{2v}=\frac{C-A}{mR^2},\]
where  $A$ is the moment
of inertia along an axis passing through  the equator and
$J_{2v}$ is the dynamic form  factor normalized by the volumetric mean
radius\footnote{The usual dynamic form  factor $J_{2}=(C-A)/mR_e^2$
  is normalized by the equatorial
  radius $R_e$. The difference  $(R_e-R)/R=\ep/3$ is of the order of
  the ellipticity $\ep$ of the body  and
  $J_{2v}=J_2\, R^2_e/R^2.$\label{footJ2}},   $\gm$ can be written as
  \begin{equation}
    \gm=\frac{\Io}{mR^2}\frac{\Om^2}{J_{2v}}
    \label{gmJ}
  \end{equation}
  The constant $\gm$ is also related to the static fluid Love number $k_2$ 
  (\cite{rr2017} equation (14))
  \begin{equation}
    k_2= \frac{3\Io G}{R^5} \frac{1}{\gm}=
    \frac{3}{2}\left(\frac{R_\Irm}{R}\right)^5\frac{\gm_\Irm}{\gm}\label{gmk2}
  \end{equation}

  The  radius $R$,
  which neither 
  appears in the definition of $\gm$ nor in the dynamic model presented in
  \cite{rr2017}, is an  important quantity in this paper.
  If
  {\it in the non-rotating state, the 
 radially symmetric  density $\rho$  of the body 
 is a non-increasing function of the radius, a hypothesis
assumed throughout the paper},  then the radius and the inertial radius
satisfy  
  \begin{equation}
 \left(\frac{R_\Irm}{R}\right)^2=\frac{5}{2}\frac{\Io}{mR^2}\le 1\label{RRIin0}
 \end{equation}
 with   $R_\Irm/R= 1$ if, and only if, the body is homogeneous
 (this well known inequality
is proved in Proposition \ref{RRIin} in Appendix
\ref{app0})).  
\footnote{If the density non-increasing hypothesis is not assumed, then   
  $R_\Irm/R\le 5/3$, where the
   equality holds for a spherical shell of radius $R$.}

The paper is organized as follows.
Sect.  \ref{clairaut} contains a  review of the theory
of Clairaut which describes the flattening of a rotating body with radial
stratification of density. This Section is subdivided into several
subsections where we discuss different types of radial density distributions:
piecewise constant, 
with a point-mass at the center,
and induced by a polytropic gas.
At the end of Section \ref{clairaut} we present the approximate theory
of Darwin-Radau. This Section does not contain any new result but  another
way to look at old formulas which highlight some of their interesting features.
For instance, the Darwin-Radau approximation
gives rise to an estimate of $\gm$ (equation (\ref{gm3})), denoted as $\gm_{DR}$, 
that depends only on the ratio
$R_\Irm/R\le 1$ and such that
\begin{equation}
  \frac{\gm_{DR}}{\gm_\Irm}=1+ \ \text{Term of order}\
    \left(1-\frac{R_\Irm}{R}\right)^3
\label{O3}
\end{equation}
Since  the Darwin-Radau approximation is exact for $R_\Irm/R=1$
we conclude that $\gm\approx \gm_\Irm$
  whenever  $R_\Irm/R\approx 1$. A plot
(Figure \ref{fradau} (a)) shows that the approximation $\gm\approx \gm_\Irm$
almost coincide with that
by the Darwin-Radau equation for  $0.89<R_\Irm/R\le 1$. For a polytrope,
$\gm\approx \gm_\Irm$ is correct up  to $3\%$ within the range
$0.7<R_\Irm/R\le 1$ (Figure \ref{fradau} (b)).

In Section \ref{secvargm} we present a variational principle for $\gm$.
There are at least two different variational  characterizations of Clairaut's equation:
one due to Macke and Voss (see \cite{moritz} Section 3.3)
and another due to Rau \cite{rau1974variational}. The variational
principle of Rau uses the adjoint equation to Clairaut's equation and is very
different from  the variational principle of Macke and Voss. 
We also present a variational
characterization of Clairaut's equation,  which is different but
related to that of
Macke and Voss,  with the additional property that the value of the
functional  to be minimized has  $\gm$ as its minimal value.
Although this variational principle can be used to estimate the value of $\gm$,
in this paper, its   relevance is mostly conceptual.

In Section \ref{riccati} we derive a new first order differential
equation that allows for the determination of $\gm$ without having to solve
Clairaut's equation. Using this equation we were able to show that
for any non increasing density function
\begin{equation}
  \frac{3}{2}\, \left(\frac{R_\Irm}{R}\right)^5\,
  \left[\frac{5}{3}\left(\frac{R}{R_\Irm}\right)^2-1\right]\le
\frac{\gm}{\gm_\Irm}
    \le \sqrt{\frac{35}{39}}\,\frac{8575}{8112}\approx 1.001401
  \label{mainin}
  \end{equation}
  This is the main  result in the paper. 
  Inequality (\ref{mainin})  implies that the Darwin-Radau theory cannot be valid if
  $\frac{R_\Irm}{R}<0.86534\ldots$
  (see the discussion close to equation \ref{limaradau}).
  If
  \begin{equation}
    \zeta=\frac{\Om^2R^3}{Gm}\label{mu}
  \end{equation}  
  denotes  the ratio of the centrifugal acceleration at the equator to the
  gravitational acceleration on the body   surface and
 $\ep(R)$ denotes the ellipticity of  the rotating figure of equilibrium, then
 inequalities (\ref{mainin}), equation (\ref{gmJ}),
 and the relation (\cite{zharkov} equation (32.19))
 \[
   3J_{2v}=2\ep(R)-\zeta
 \]
 imply
 \begin{equation}
   2\bigg(1-\frac{3}{5}\left(\frac{R_\Irm}{R}\right)^2\bigg)\le
   \frac{\zeta}{\ep(R)}\le 2
   \bigg(1+1.4979\left(\frac{R_\Irm}{R}\right)^5\bigg)^{-1}
\label{epr}
\end{equation}
or, equivalently,
\begin{equation}
  \frac{1}{3}\left(2-\frac{\zeta}{\ep(R)}\right)\le
  \frac{\Io}{mR^2}\le
  0.400225\,\left(\frac{4}{3}\,\frac{\ep(R)}{\zeta}-\frac{2}{3}\right)^{2/5},
    \label{IomR2}
  \end{equation}
  or
  \begin{equation}
    2\left(\frac{\zeta}{J_{2v}}+3\right)^{-1}\le \frac{\Io}{mR^2}\le
    0.400225\left(\frac{\zeta}{2J_{2v}}\right)^{-2/5}.
      \label{IomR22}
\end{equation}  
For a homogeneous body $\zeta/\ep(R)=4/5$ and $\zeta/J_{2v}=2$ and the above
inequalities
 are almost sharp, namely  $0.4\le \frac{\Io}{mR^2}\le  0.400225$. 

Inequalities \ref{epr} must be compared to those of Chandrasekhar
\cite{chandra1933}, \cite{chandra1963}
\[
\frac{4}{5}\le \frac{\zeta}{\ep(R)}<2
\]
obtained without any constraint on the ratio $R_\Irm/R$
(according to \cite{chandra1933}, equation (100), the first lower bound was known to
Laplace).
The lower bound in inequality  (\ref{epr}) coincides with the lower bound
by Chandrasekhar if $R_\Irm/R=1$ and the upper bound in inequality 
(\ref{epr}) coincides with the upper bound
by Chandrasekhar if $R_\Irm/R=0$. For other values of $R_\Irm/R$ our inequalities,
which essentially cannot be improved without further constraining the density, 
represent a great advance with respect to those of Chandrasekhar.

In Section \ref{computation} we present an application
of the results in the previous sections to  large bodies
in the Solar system (Sun, Earth, Mars, Jupiter, Saturn, Uranus and Neptune).
The value of $\gm$ is estimated by different means:
 integration of Clairaut's equation using  density functions 
available in the literature, the Darwin-Radau approximation, and
 equation (\ref{gmJ}) with  values
of $\Om$ and $J_{2v}$ obtained from observations.
The results are summarized in tables and figures given in Section \ref{computation}.

Finally, Section \ref{conc} is a conclusion where we propose  a 
way to estimate   $\gm$ 
in terms of  $m$, $\Io$, and $R$.
Polytropes are taken as archetypal models and it is verified 
that the values of $\gm$ obtained from  this model
are close to those estimated from observations.

The results in this paper require careful interpretation. Unless a body
is homogeneous, for which $R_\Irm/R=1$ and $\gm/\gm_\Irm=1$,
there is no one-to-one relation between the normalized
gravitational  modulus $\gm/\gm_\Irm$ and the ratio $R_\Irm/R$.\footnote{For
  given values of $m$,
  $\Io$, and $R$ such that $R_\Irm/R<1$
    there is an internal density distribution that realizes the lower bound in
    inequality
    (\ref{mainin}),
    the ``Thick shell Roche model'', and another that realizes $\gm/\gm_\Irm=1$, 
    the ``Homogeneous core Roche model''
    (see Section \ref{roche} for details).
    There is always
    an internal density that realizes  an intermediate value of $\gm/\gm_\Irm$.
  }
  Nevertheless, the  remarkable Darwin-Radau theory and the results in
  this paper show that, for most large bodies in the Solar system,
  $\gm$ is  essentially determined by $m$, $\Io$, and $R$, and
  for a body with $0.7 \le R_\Irm/R <1$ the value of $\gm$ is close to
  that of an equivalent
  uniform body with radius $R_\Irm$.  By no means the internal density
  $\rho$ of the body of radius $R$, which must satisfy $\rho(r)>0$ for $R_\Irm<r<R$,
  has to be close to that of the uniform body of radius $R_\Irm$. Moreover,
  the empirical conclusion of the paper is that, for a large  body,
  if $m$, $\Io$, and $R$ are known  then
   its gravitational modulus $\gm$
  is approximately that of a polytrope with the same values of the known quantities
  (for $0.7 \le R_\Irm/R <1$
  this approximation is
  even better than $\gm\approx\gm_\Irm$).
  Again, by no means the density of the given body must be close to that of a
  polytrope.

  This work has been done under the premise that $\Io$ is given
  but ``this  is exactly the opposite to the actual practice followed by
  astronomers'' (referee's comment)  who from observed values of  $\Om$ and several
     gravitational moments $J_2,J_4,J_6,\ldots$ obtain information about  
     the internal mass distribution of the body including the value of $\Io$
     (see \cite{zharkov} chapter 3). The next paragraph shows that in situations
     where astronomical observations are difficult the results in this paper
     can be used in the usual astronomer way.

The spin rate of an exoplanet can be  estimated 
from near-infrared spectroscopic observations \cite{snellen2014fast}.
As argued in \cite{kellermann2018interior},
``the shape of the transiting light curve
might, in principle, reveal the shape of the planet, and in particular,
its deviation from spherical symmetry'', namely $\ep(R)$.
So, in theory, it may be possible to estimate $\zeta$ and $\ep(R)$
for an exoplanet and, using    $3J_{2v}=2\ep(R)-\zeta$, also
$J_{2v}$. If this is the case, then  equation
(\ref{IomR2}) establishes  upper and lower bounds to $\Io/(mR^2)$. 
If the approximation $\gm\approx\gm_\Irm$ is further adopted
then equations (\ref{gmI}), (\ref{gmJ}),  (\ref{RRIin0}), and (\ref{mu}) imply 
\begin{equation}
 \frac{\Io}{mR^2}\approx
 0.4\,\left(
 \frac{2 J_{2v}}{\zeta}\right)^{2/5},
  \label{epmuapprox}
\end{equation}  
which is a as Darwin-Radau type formula.
The approximation may fail: for bodies with unusual density
 distributions (as for instance with a high density core
 surrounded by a homogeneous layer, see
 the ``Thick shell Roche model'' in Section \ref{roche}), for fast spinning bodies
 (corrections of order $\zeta^2$ are nonnegligible), and for  bodies that
 are not in hydrostatic
 equilibrium.

In order to test the validity of equation (\ref{epmuapprox}),
 the values of $m$, $R$, $\ep(R)$, and $\Om$ for
 the Earth, Mars, Jupiter, Saturn, Uranus, and Neptune,
  taken from the ``NASA fact sheets'', were used to estimate
  $J_{2v}=(2\ep(R)-\zeta)/3$ and, by means of equation (\ref{epmuapprox}),
  $\frac{\Io}{mR^2}$. The result was compared to the values of $\frac{\Io}{mR^2}$
  given in the same reference. The relative difference between both values
  varies from $0.7\%$ (for the Earth) to $20\%$ (for Uranus) being in the average
  $10\%$. Notice that within Clairaut's theory
  the approximation $\gm\approx\gm_\Irm$ for
  the gravitational modulus $\gm$ is much better  than that for $\frac{\Io}{mR^2}$
  given in equation (\ref{epmuapprox}).
  This may be explained by the variational principle: $\gm$ is the minimum  value
  of a functional that has as its minimum  point the
  the  ellipticity (times an arbitrary nonnull constant) as a function of the radius.
  An error of order $\dt$ in  the minimum point  becomes an
  error of order $\dt^2$ in the minimum value $\gm$.

\section{Clairaut's Equation}

\label{clairaut}

The main goal in this section is to review some results about  Clairaut's
equation and to solve it in some special cases. The theory of figures of Clairaut
is explained in detail in \cite{poincare1902figures}, \cite{moritz},
and  \cite{zharkov}. 

The Clairaut's equation (first obtained in 1743) describes the equilibrium
configuration of a spherically symmetric
self-gravitating celestial body made of an  incompressible fluid.
The body is supposed to be
rotating steadily with uniform angular speed $\Om$ about a fixed axis $\erm_3$ passing
through its center of mass.
In the rotating frame the configuration 
must satisfy the stationary Euler's equation  given by
\begin{equation}
\frac{1}{\tilde \rho}\nabla p=-\nabla [\Phi+\Phi_c]
\label{eu2}
\end{equation}
where $x$ is the position vector, 
$\tilde \rho(x)$ is the density, $p(x)$ is the pressure,
$\Phi_c=-\frac{\Om^2}{2}(x_1^2+x_2^2)$
  is the potential of the centrifugal force
and $\Phi$ is the gravitational potential  given by
\begin{equation}
  \Phi(x)=
  -G\int_{\R^3}\frac{\tilde\rho(\tilde x)}{|x-\tilde x|} d\tilde x\label{grav1}
\end{equation}    
Equation (\ref{eu2})
shows that $\nabla \tilde \rho\times \nabla [\Phi+\Phi_c]=0$, so the level sets of all three functions $\tilde \rho$, $\Phi+\Phi_c$, and $p$
  coincide.

  For  $\Om=0$
  equation (\ref{eu2}) has a solution with spherical symmetry.
  For $\Om>0$ sufficiently small we may  expect
  the existence of solutions  with  level sets which are approximately
  ellipsoids of revolution. More precisely, if for $\Om=0$ the radius of
  a spherical shell of constant density is $r>0$, then for $\Om>0$
  this shell becomes ellipsoidal and is given by
\begin{equation}
\frac{x_1^2+x_2^2}{r^2(1+\ep/3)^2}+\frac{x_3^2}{r^2(1-2\ep/3)^2}=1
\label{ell}
\end{equation}
where $\ep(r)>0$ is the small  flattening of the ellipsoid, defined as the ratio
(equatorial radius - polar radius)/(equatorial radius).

In spherical coordinates $(r,\theta,\phi)$,   with polar axis given by $\erm_3$,
it can be shown that (see equation (\ref{phi12}) in Appendix \ref{app1}) 
\begin{equation}
  \begin{split}
    & \Phi(r,\theta)=\Phi_0(r)+\Phi_2(r)P_2(\cos\theta)+\Oc(\ep^2)\qquad\text{where:}\\
&P_2(\cos\theta)=\frac{1}{2}(3\cos^2\theta-1),\\
    & \Phi_0(r)=  -4\pi G\left\{\frac{1}{r}\int_{0}^r a^2 \rho(a)da
  +\int_{r}^\infty a \rho(a)da\right\},\\
  & \Phi_2(r)=  -\frac{8\pi}{15} G\left\{\frac{1}{r^3}
  \int_{0}^r a^5\ep(a) \rho^\prime(a)da
  +r^2\int_{r}^\infty \ep(a) \rho^\prime(a)da\right\},
    \end{split}\label{phi1}
\end{equation}
$\rho$  is the spherically symmetric density function of the body at rest
with $\rho(r)=0$ if $r>R$, and a prime denotes a derivative with respect to $r$.
The  gravitational potential external to the body 
can also be written  as
  \begin{equation}
    \Phi(r,\theta) = - \frac{G\,m}{r} +J_{2v}\,\frac{G\,m\,R^{\,2}}{r^{\,3}}\,
    P_2(\cos\theta)+\Oc(r^{-4}).
  \label{phirtheta}
  \end{equation}
  where
   \begin{equation}
    J_{2v}=-\frac{8\pi}{15} \frac{1}{mR^2}  \int_{0}^\infty a^5\ep(a) \rho^\prime(a)da.
\label{J2}
\end{equation}

Suppose that 
\begin{equation}
  \Om^2=\Oc(\ep) \quad \text{and}\quad
\zeta=\frac{\Om^2R^3}{Gm}\ \ \text{are small}.
\label{omep} 
\end{equation}
Under these hypotheses a nontrivial argument,
where terms of order $\ep^2$ are neglected
(see, for instance, \cite{cook2009interiors}, \cite{moritz}, or \cite{zharkov}), 
shows that 
\[
-\frac{2}{3}r\ep(r)\Phi_0^\prime(r)+\Phi_2(r)+
  \frac{\Om^2r^2}{3}=0
  \]
This is Clairaut's integral equation that  can  be written as 
\begin{equation}
  4\pi G\left\{\ep(r) r^2\int_{0}^r a^2\rho(a)da
+\frac{1}{5} \int_{0}^r a^5\ep(a) \rho^\prime(a)da
+\frac{r^5}{5}\int_{r}^\infty \ep(a) \rho^\prime(a)da\right\}=
\frac{\Om^2 r^5}{2}.\label{cl1}
\end{equation}

Clairaut's equation can be presented in different forms, which are
interesting for different reasons. 
Let 
\begin{equation}
  \ov\rho (r)=\frac{3}{r^3}\int_0^ra^2\rho(a)da
  \label{ovrho1}
\end{equation}
be the mean density inside the spheroidal shell of radius $r$ and
\begin{equation}
\begin{split}
  K(a,r)&=
\rho^\prime(a)\rho^\prime(r)F(a,r), \quad\text{where}\\
F(a,r)&=  \begin{cases} 
    \frac{r^5}{5} & \text{if}\quad a\ge r \\
    \frac{a^5}{5} & \text{if}\quad r\ge a, 
   \end{cases}
    \end{split}
  \label{K}
\end{equation}  
be a positive symmetric function. 
Note that
\begin{equation}
\ov\rho^\prime(r)=\frac{3}{r}\bigl[\rho(r)-\ov\rho(r)\bigr].
\label{ovrho2}
\end{equation}
Equation (\ref{cl1}) multiplied by $\rho^\prime$ can be written as
\begin{equation}
  4\pi G\left\{\frac{r^5\ov\rho(r)\rho^\prime(r)}{3}\ep(r)
  +\int_{0}^\infty K(a,r)\ep(a) da\right\}=\Om^2\frac{r^5 \rho^\prime(r)}{2},\label{cl2}
\end{equation}
where we used
\begin{equation}
\int_{0}^\infty K(a,r)\ep(a) da=\rho^\prime(r)
\left\{\frac{1}{5} \int_{0}^r a^5\ep(a) \rho^\prime(a)da
+\frac{r^5}{5}\int_{r}^\infty \ep(a) \rho^\prime(a)da\right\}
\label{Kaux}
\end{equation}

Integration by parts and   differentiation of equation
(\ref{cl1}) with respect to $r$ gives
\begin{equation}
  \left(r\,\epsilon^\prime+2\,\epsilon\right)\ov{\rho}(r) -
  3\int_{r}^\infty \ep^\prime (a) \rho(a)da=\frac{15\,{\Omega}^{\,2}}{8\pi\,G}.
  \label{cl3}
  \end{equation}
Further differentiation with respect to $r$ gives 
the Clairaut's differential equation
\begin{equation}
  r^{\,2}\,\ep^{\, \prime\prime}-6\,\epsilon + 6\,\frac{\rho}{\ov{\rho}}
  \left(r\,\epsilon^{\,\prime}+\epsilon\right)= 0
  \label{cl4}
 \end{equation} 
that can also be written as 
\begin{equation}
r\,\ep^{\, \prime\prime} +  6\,\epsilon^{\,\prime}+
2\,\frac{\ov\rho^\prime}{\ov{\rho}}
  \left(r\,\epsilon^{\,\prime}+\epsilon\right)= 0,
  \label{cl5}
\end{equation}
where we used equation (\ref{ovrho2}).

 Many realistic models for $\rho$
have points of discontinuity as, for instance, 
the ``Preliminary Reference Earth Model'' (PREM) \cite{prem}. In the following
we rewrite Clairaut's  equation (\ref{cl1}) in a way
which is convenient for working with  
densities that satisfy the following hypothesis:
  \begin{equation}
  \begin{split}
    &\rho  \ \text{is: non-increasing;}
    \   \text{piecewise}\ C^2 \ \text{with finitely many
  points of}\\ &\text{discontinuities} \ 0<r_1<r_2,\dots,<r_n\le R;
  \ \text {and} \  \rho(r)=0\ \text{for}\ r>R.
  \end{split}
  \label{innerrho2}
  \end{equation}  
Let
 \begin{equation}
   \begin{split}
     q(r)&= \frac{1}{5}\int_r^\infty \rho^\prime(a)\ep(a)da\\
     \sg(r)&=\frac{1}{5r^5}\int_0^ra^5\rho^\prime(a)\ep(a)da+q(r),
   \end{split}
   \label{q0}
 \end{equation}
which together  with equation (\ref{phi1}) imply 
 \begin{equation}
 \Phi_2(r)= -\frac{8\pi}{3} G\, r^2\sg(r).
 \label{phisg0}
 \end{equation}
 If $\rho$ is $C^2$ then Clairaut's equation (\ref{cl1}) can be written as the
 following boundary
 value problem
 \begin{equation}
   \begin{split}
     q^\prime&=-\frac{1}{5}  \rho^\prime\ep\\
     \sg^\prime&=-\frac{5}{r}(\sg-q)\\
     \ep(r)&=\frac{3}{\ov\rho(r)}\left(\frac{\Om^2}{8\pi G}-\sg(r)\right)\\
     q(0)&=\sg(0)=q_0,\quad\text{such that}\quad q(R)=0,
   \end{split}
   \label{cl6}
 \end{equation}
 where $q(0)=\sg(0)=q_0$ must be understood as $\lim_{r\to 0}[q(r)-\sg(r)]=0$ with $r>0$.
 If $\rho$ has points of discontinuity as in hypothesis (\ref{innerrho2}),
then let $\chi_j$ be the density jump at $r_j$,
\begin{equation}
  \chi_j=\rho(r_j)-\lim_{r\to (r_{j})_-} \rho(r)< 0\quad\text{for}\quad j=1,\ldots n,
  \label{rhod}
  \end{equation}
where $r\to (r_{j})_-$ denotes the limit as $r$ tends to $r_j$ with
$r<r_j$. 
The derivative of $\rho$ at $r_j$, in distribution sense,  is
$\rho^\prime(r)=\chi_j\dt(r-r_j)$ where $\dt$
is the Dirac $\dt$-measure. Using this fact and integrating equations (\ref{cl6})
in a neighborhood of $r_j$ we obtain the following jump conditions at $r_j$: 
\begin{equation}
   \begin{split}
     \Delta q(r_j)&= q(r_j)-\lim_{r\to (r_{j})_-} q(r)=-\frac{\chi_j}{5}\ep(r_j)\\
     \Delta \sg(r_j)&= \sg(r_j)-\lim_{r\to (r_{j})_-} \sg(r)=0 
   \end{split}
   \label{jumpcl}
\end{equation}
Equations (\ref{cl6}) plus the jump conditions (\ref{jumpcl}) entirely determine
the solution to Clairaut's equation (\ref{cl1}).
We remark that $\sg$ and  $\ep$ are continuous functions
even when $\rho$ is not and that if
$r_n=R$ is a point of discontinuity,  then $q(R)=0$ in equation (\ref{cl6})
must be understood as
\begin{equation}
  0=q(R)=\Delta q(r_n)+\lim_{r\to (r_{j})_-} q(r)=-\frac{\chi_n}{5}\ep(r_n)
 + \lim_{r\to (r_{j})_-} q(r)
  \label{qR0}
  \end{equation}
Equations  (\ref{phisg0}), (\ref{cl6}), and 
(\ref{J2}) imply 
 \begin{equation}
   J_{2v}=-2\frac{\sg(R)}{\ov \rho(R)}=-\frac{8\pi R^3}{3m}\sg(R).
   \label{J5}
 \end{equation}
 and, the well known relation,
   \begin{equation}
    J_{2v}=\frac{1}{3}\left(2\ep(R)-\frac{\Om^2R^3}{Gm}\right),
    \label{J3}
    \end{equation}

In order to solve the boundary value problem in equation (\ref{cl6})
it is convenient to  further change of variables as 
 \begin{equation}
 w=q-\frac{\Om^2}{8\pi G},\quad y=\sg-\frac{\Om^2}{8\pi G} \label{trans1}
\end{equation}
Then equation (\ref{cl6}) becomes 
 \begin{equation}
   \begin{split}
     w^\prime&=\frac{3}{5}\frac{\rho^\prime}{\ov \rho} y\\
     y^\prime&=-\frac{5}{r}(y-w)\\
     \ep(r)&=-\frac{3}{\ov\rho(r)}y,
   \end{split}
   \label{cl8}
 \end{equation}
 the boundary conditions become
 \begin{equation}
        w(0)=y(0)=w_0,\quad w(R)=-\frac{\Om^2}{8\pi G},\label{bc8}
 \end{equation}
 and the jump conditions become
\begin{equation}
   \begin{split}
     \Delta w(r_j)&=\frac{3}{5}\frac{\chi_j}{\ov\rho(r_j)}y(r_j)\\
     \Delta y(r_j)&=0.
   \end{split}
   \label{jumpcl2}
\end{equation}

The  following Proposition is well known in the case where $\rho$
is $C^2$
(see, for instance,
\cite{poincare1902figures} chapter IV)
and due to its importance in this paper it is proved  in  Appendix \ref{app0}.
  \begin{proposition} Suppose that $\rho$ satisfies hypothesis {\rm (\ref{innerrho2})}.
    Then, for $\Om>0$  there exists a unique bounded solution to equation
    {\rm  (\ref{cl1})} {\rm (}and therefore to problems {\rm  (\ref{cl6})} and
    {\rm  (\ref{cl8})}{\rm )}. This solution is strictly positive, non-decreasing,
    and     $C^1$.
     For $\Om=0$
    the only solution to equation {\rm  (\ref{cl1})}
     is $\ep(r)=0$,  $r\ge 0$.
  \label{stat1}
   \end{proposition}

 The following algorithm allows for the solution to equation (\ref{cl1}). 
 Let $(\tilde w, \tilde y)$ be the solution to the differential equation
 (\ref{cl8}) with the initial condition $\tilde w(0)=\tilde y(0)=1$ and
  the jump conditions (\ref{jumpcl2}). We claim that
  $\tilde w(R)\ne 0$. Indeed, if $\tilde w(R)= 0$ then 
  $(\tilde y, \tilde w)$ satisfies the boundary conditions (\ref{bc8})
  with $\Om=0$ and the corresponding  $\ep$ solves equation
  (\ref{cl1}) with $\Om=0$, which is impossible due to Proposition \ref{stat1}.
 The desired solution
 $(w, y)$ to the boundary value problem in equations
 (\ref{cl8}), (\ref{bc8}), and (\ref{jumpcl2})
   satisfies the initial condition
 \begin{equation}
 w(0)=y(0)=-\frac{\Om^2}{8\pi G}\frac{1}{\tilde w(R)}
 \label{trans2}
 \end{equation}

 \subsection{N-layer models}
 
\subsubsection{Piecewise constant density functions}
\label{piecewise}

In this paragraph we consider functions $\rho$ of the following form
\begin{equation}
  \rho(r)=\left\{
 \begin{array}{lll}
   \rho_0=\text{constant}>0&\text{for}&  0=r_0\le r <r_1\\
   \rho_1=\text{constant}>0&\text{for}&  r_1\le r <r_2\\
   \cdots & & \\
   \rho_{n-1}=\text{constant}>0&\text{for}&  r_{n-1}\le r <r_n=R\\
   \rho_{n}=0&\text{for}&  R\le r
 \end{array}
 \right.\label{rhoconst}
 \end{equation}
In this case the  solution to equation (\ref{cl6}) is continuously
differentiable in each
interval $r_{j-1}\le r<r_j$, $j=1,\ldots n$. Using the definitions
\[
  q_j= q(r_j),\quad
  \sg_j= \sg(r_j),\quad     \ep_j= \ep(r_j),\quad j=0,\ldots n,
  \]
and the initial conditions   $q(0)=\sg(0)=q_0$,
we can write   the solution to equation (\ref{cl6}) 
  as
\begin{equation}
  q(r)= q_0,\quad  \sg(r)= q_0,\quad \ep(r)=\frac{3}{\rho_0}
  \left(\frac{\Om^2}{8\pi G}-q_0\right)
  \text{for}\quad 0\le r<r_1 \label{solcl0}
\end{equation}
The jump  conditions at equation (\ref{jumpcl}) imply
\begin{equation}
  \begin{split}
     q_1&= q_{0}-\frac{\rho_1-\rho_{0}}{5}\ep_1\\
     \sg_1&=q_0\\
  \ep_1&=\frac{3}{\rho_0}
  \left(\frac{\Om^2}{8\pi G}-q_0\right)
\end{split}
  \label{solcl01}
\end{equation}  
For $r_{j}\le r <r_{j+1}$, $j=1,\ldots n-1$, the solution
to equation (\ref{cl6}) 
can be written as
\begin{equation}
 \begin{split}
     q(r)&= q_j\\
     \sg(r)&= \sg_j\frac{r_j^5}{r^5}+q_j\left(1-\frac{r_j^5}{r^5}\right)\\
     \ep(r)&=\frac{3}{\ov\rho(r)}\left(\frac{\Om^2}{8\pi G}-\sg(r)\right)
\end{split}
\label{solcl1}
\end{equation}
 Using  the jump conditions in equation (\ref{jumpcl})
we obtain that, for $j=2,\ldots n$, 
\begin{equation}
   \begin{split}
     q_j&= q_{j-1}-\frac{\rho_j-\rho_{j-1}}{5}\ep_j\\
     \sg_j&=\sg_{j-1}\frac{r_{j-1}^5}{r_j^5}+q_{j-1}\left(1-\frac{r_{j-1}^5}{r_j^5}\right)\\
     \ep_j&=\frac{3}{\ov\rho(r_j)}\left(\frac{\Om^2}{8\pi G}-\sg_j\right)
   \end{split}
   \label{solcl2}
\end{equation}
For numerical computations it is convenient to add the following recursion relation for
$\ov\rho(r_j)\defi\ov\rho_j$:
\begin{equation}
  \ov\rho_j=\rho_{j-1}+\frac{r_{j-1}^3}{r_j^3}(\ov\rho_{j-1}-\rho_{j-1}),\quad j=1,\ldots,n,\quad\text{with}\quad
  \ov \rho_0=\rho_0
  \label{ovrhoj}
  \end{equation}
Finally, the boundary condition $q(R)=0$ implies 
\begin{equation}
 q_n=0
  \label{bccl6}
\end{equation}
and equation (\ref{J5}) implies
 \begin{equation}
   J_{2v}=-2\frac{\sg_n}{\ov \rho_n}=-\frac{8\pi R^3}{3m}\sg_n \label{J6}
 \end{equation}

 The solution to this boundary value problem can be easily obtained
 using the transformation (\ref{trans1}) and the initial condition (\ref{trans2}).
 The method just presented is well known in  geophysics as the ``propagation
 matrix method''.
 
 \subsubsection{Two-layer models and generalized  ``Roche models''} 
\label{roche}
 
Consider   a density function with just two different values
with   $x=r_1/R$ denoting the internal point of discontinuity.
Then, after using  a computer algebra system,
we obtain
\begin{equation}
  J_{2v}=-2\frac{\sg_2}{\ov \rho(R)}=\frac{\Om^2R^3}{Gm}\,
  \frac{\rho_1(2\rho_0 + 3\rho_1) + (\rho_0 - \rho_1)(5\ov\rho(R) + 3\rho_1)x^5}
{(5\ov\rho(R) - 3\rho_1)(2\rho_0 +3\rho_1) - 9\rho_1(\rho_0 -\rho_1)x^5}
\label{J7}
\end{equation}
This expression,  equation (\ref{gmJ}) and the definition
$\gm_\Irm=(4/5)(Gm/R_\Irm^3)$ imply
\begin{equation}
  \frac{\gm}{\gm_\Irm}=\frac{1}{2}\, \left(\frac{R_\Irm}{R}\right)^5
\frac{(5\ov\rho(R) - 3\rho_1)(2\rho_0 +3\rho_1) - 9\rho_1(\rho_0 -\rho_1)x^5}
     {\rho_1(2\rho_0 + 3\rho_1) + (\rho_0 - \rho_1)(5\ov\rho(R) + 3\rho_1)x^5}
\label{gm4}     
\end{equation}
This expression can be further simplified using the relations
\[
\ap=\frac{\rho_1}{\rho_0}\in(0,1),\quad \text{ and} \quad
\ov\rho(R)=\rho_1+x^3(\rho_0-\rho_1),
\]
the result is
\begin{equation}
  \frac{\gm}{\gm_\Irm}=\frac{1}{2}\, \left(\frac{R_\Irm}{R}\right)^5
  \frac{3 \ap^2 \left(3 x^5-5 x^3+2\right)+\ap \left(-9 x^5+5 x^3+4\right)+10 x^3}
       {\ap^2 \left(5 x^8-8
    x^5+3\right)+\ap \left(-10 x^8+8 x^5+2\right)+5 x^8}
\label{gm5}     
\end{equation}
where 
\begin{equation}
  \frac{R_\Irm}{R}=\sqrt{\frac{x^5+\ap(1- x^5)}{x^3+\ap(1- x^3)}}
  \label{RIR}
  \end{equation}
or
\begin{equation}
  \ap=\frac{\left(\frac{R_\Irm}{R}\right)^2x^3-x^5}
     {\left(\frac{R_\Irm}{R}\right)^2x^3-x^5+1-\left(\frac{R_\Irm}{R}\right)^2}
\label{ap}
\end{equation}
Notice that the condition $0<\ap\le 1$ and equation (\ref{ap}) imply that
\begin{equation}
  0<x\le\frac{R_I}{R}
  \label{inx}
  \end{equation}

  Two interesting limits of two layer bodies are discussed in the following. 
  These limits are generalizations of the
  usual Roche model, which consists of  a point mass
  surrounded by a medium so rarefied that its mass can be neglected.

  \nd {\it Homogeneous core Roche model:} is the  body
  obtained as $\ap\to 0$ while $x$ remains fixed. So the limit body is just
  a homogeneous body of density $\rho_0$ and radius $R_\Irm$
  surrounded by a rarefied layer of thickness $R-R_\Irm$ of negligible density.
  The family of homogeneous core Roche models is parameterized by
  $x=R_\Irm/R\in (0,1)$  and, for all $x$, $\gm/\gm_\Irm=1$ in agreement with equations
  (\ref{RIR}) and (\ref{gm5}).
  The fact that  for all $x$, $\gm/\gm_\Irm = 1$,
  follows from  our definition of rigidity that is 
  associated with the dynamical flattening $(C-\Io)/\Io$  and not to the geometrical one
$(a-c)/a$,  where $(a, b, c)$ are the semi-axes of the ellipsoidal shape of the body.
Therefore, the massless atmosphere has
no dynamical effect.

 \nd {\it Thick shell Roche model:}
  The second limit is more interesting and occurs as $x\to 0$
  while $R_\Irm/R$ remains fixed. In this case
  equation (\ref{ap}) implies
  \begin{equation}
  \ap=\frac{\left(\frac{R_\Irm}{R}\right)^2}
           {1-\left(\frac{R_\Irm}{R}\right)^2}\, x^3+\Oc(x^5)
           \label{aplim}
  \end{equation}         
and equation (\ref{gm5}) implies
\begin{equation}
  \lim_{x\to 0}
  \frac{\gm}{\gm_\Irm}=
  \frac{1}{2}\, \left(\frac{R_\Irm}{R}\right)^3\,
  \left[5-3\left(\frac{R_\Irm}{R}\right)^2\right]
\label{limgm}     
\end{equation}
The equation for the moment of inertia,
$\Io=\frac{8\pi}{3}\int_0^{R} r^4\rho(r)ds$, implies
\[
\ov\rho(R)\left(\frac{R_\Irm}{R}\right)^2=\rho_1+x^5(\rho_0-\rho_1),
\]
and from this equation and equation (\ref{aplim}) follow 
\[
\rho_1\to \ov\rho(R)\left(\frac{R_\Irm}{R}\right)^2\quad\text{as}\quad x\to 0 
\] 
and
\[
\rho_0=\frac{\rho_1}{\ap}=\frac{\ov\rho(R)}{x^3}
\left[1-\left(\frac{R_\Irm}{R}\right)^2\right]+\Oc(x^{-1})
\]
So, if $R_\Irm/R<1$,  the limit body obtained as $x\to 0$ represents
 a  family of ``Roche models''
that consists of a 
point at the origin with mass $m_0=[1-(R_\Irm/R)^2]m$ and a surrounding
homogeneous layer with  mass $m_1=(R_\Irm/R)^2m$.
 The sum $m_0+m_1=m$ gives the
total mass of the body and the moment of inertia  is only 
due to  the homogeneous layer with $\Io=0.4\, m_1R^2=0.4\, mR_\Irm^2$.
Notice that  the limit as $R_\Irm/R\to 0$ represents the actual  Roche model where 
the surrounding medium is so rarefied that the  moment of inertia of the body
can be neglected.

\subsection{Polytropic models}
\label{polysec}

\begin{figure}[t]
\begin{center}
\includegraphics[width=\textwidth]{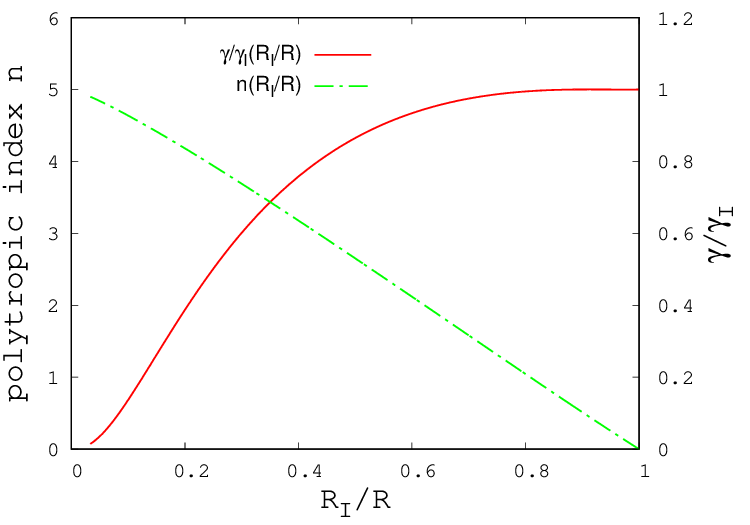}
\end{center}
\caption{  
  Graph of the polytropic index $n$ of a polytrope
  as a function of $R_\Irm/R$ (left vertical axis)
  and graph of the ratio $\gm/\gm_\Irm$ of a polytrope  as a function of
$R_\Irm/R$ (right vertical axis).
} 
\label{figpolr}
\end{figure}

The interior density distribution of stars and fluid planets can be easily determined
under the hypothesis that they are made of a polytropic fluid such that 
 pressure depends upon  density as 
 \begin{equation}
 p=K\rho^{1+1/n}\label{pol1}
 \end{equation}
 where $K>0$ is a constant and $n$ is the polytropic index.
 For  $\Om=0$,   equation (\ref{eu2}) for the hydrostatic equilibrium
 becomes
 \begin{equation}
 \frac{p^\prime}{\rho}=-\Phi_0^\prime=-\frac{4\pi G}{r^2}\int_0^ra^2\rho(a)da
 \label{pol2}
 \end{equation}
 where we used the expression for $\Phi_0$ given in equation (\ref{phi1}).
 If we write
 \begin{equation}
 \rho(r)=\lb \theta^n(r),\label{pol3}
 \end{equation}
 where $\lb>0$, then equations (\ref{pol1}) and (\ref{pol2}) imply that $\theta$
 must satisfy
 \[
 \frac{K(n+1)\lb^{1/n-1}}{4\pi G}\frac{\big(r^2\theta^\prime\big)^\prime}{r^2}=
 -\theta^n
 \]
 If a new spatial variable $\xi$ is defined as
 \begin{equation}
   r=\left(\frac{K(n+1)\lb^{1/n-1}}{4\pi G}\right)^{1/2}\xi
   \label{pol4}
 \end{equation}
 then we obtain the so called Lane-Emden equation
 \begin{equation}
   \frac{1}{\xi^2}\frac{d}{d\xi}\left(\xi^2\frac{d\theta}{d\xi}\right)=
   \frac{\big(\xi^2\theta^\prime\big)^\prime}{\xi^2}= -\theta^n
 \label{emden}
 \end{equation}
 The initial conditions are:  $\theta(0)=1$, which is just a normalization
 such that $\lb=\rho(0)$, and $\theta^\prime(0)=0$, which is due to the
 regularity of the density at $r=0$. For $0<n<5$ the Lane-Emden equation
 has a solution $\theta$ that reaches zero at a finite value of $\xi$ denoted
 as $\xi_1$.
 So, for each value of $n\in(0,5)$ the solution to the Lane-Emden equation,
 which is refereed as a polytrope, 
 defines a body of radius
 \begin{equation}
   R=\left(\frac{K(n+1)\lb^{1/n-1}}{4\pi G}\right)^{1/2}\xi_1\Longrightarrow
   r=\frac{R}{\xi_1}\xi
   \label{pol5}
 \end{equation}
 Notice that,  given $\lb$ and $n$,  $R$ is determined by $K$.
The  mass of the body is given by 
 \begin{equation}
   m=4\pi\int_0^R\rho(r) r^2dr=
   4\pi\lb\frac{R^3}{\xi_1^3}\int_0^{\xi_1}\theta^n(\xi) \xi^2d\xi  
 \label{pol6}
 \end{equation}
 and  the moment of inertia of the body is given by
 \begin{equation}
   \Io=\frac{8\pi}{3}\int_0^R\rho(r) r^4dr=\frac{8\pi}{3}\lb\frac{R^5}{\xi_1^5}
   \int_0^{\xi_1}\theta^n(\xi) \xi^4d\xi  
 \label{pol7}
 \end{equation}
 Equations (\ref{pol6}) and (\ref{pol7}) and the relation $\Io/(mR^2)=0.4(R_\Irm/R)^2$
 imply
 \begin{equation}
   \left(\frac{R_\Irm}{R}\right)^2=\frac{5}{3}\frac{1}{\xi_1^2}
   \frac{\int_0^{\xi_1}\theta^n(\xi) \xi^4d\xi}{\int_0^{\xi_1}\theta^n(\xi) \xi^2d\xi}
   \label{polr}
 \end{equation}  
 This equation shows that $R_\Irm/R$ is determined entirely by the polytropic index,
 it neither depends on $\lb=\rho(0)$ nor on $K$. Figure \ref{figpolr},
 which was obtained from the numerical integration of the Lane-Emden equation,
 shows that the relation of $R_\Irm/R$ and $n$ is one-to-one and almost affine.

 Given a  density distribution $\rho$  determined by a polytropic
 fluid it is possible to numerically integrate Clairaut's equation  to obtain
 the ratio $\gm/\gm_\Irm$.
As said in the Introduction, $\gm$ is invariant under
homothetic transformations that preserve the density,  so $\gm$ can be computed
using a body of radius $R=\xi_1$. Moreover, 
$\gm/\gm_\Irm$ is additionally invariant under the multiplication of $\rho$
by a constant (see, for instance,  equation (\ref{gmgmI})), 
so $\gm/\gm_\Irm$ does not depend on  $\lb$ either, and
we obtain that $\gm/\gm_\Irm$ is determined entirely
by the  polytropic index $n$. Since there is a one-to-one correspondence
between $n$ and  $R_\Irm/R$, the value of $\gm/\gm_\Irm$ for bodies modeled by polytropes
  is fully  determined by the ratio $R_\Irm/R$. The graph
 of $\gm/\gm_\Irm$ as a function of $R_\Irm/R$ is shown in Figure \ref{figpolr}.

\subsection{ The Darwin-Radau approximation}
\label{darwin-radau}

The Darwin-Radau approximation consists on the substitution \cite{cook2009interiors},
\cite{moritz}, \cite{zharkov}
\[
\eta = \frac{r}{\epsilon}\,\epsilon^\prime
\]
into equation (\ref{cl5}) that leads to the equation
\begin{equation}
\frac{d}{dr}\left[\ov{\rho}\,r^{\,5}\,(1+\eta)^{1/2}\right] =
5\,\ov{\rho}\,r^{\,4}\,\psi(\eta),
\label{dr}
\end{equation}
where
\[
\psi(\eta) =\frac{1+\eta/2-\eta^{\,2}/10}{(1+\eta)^{1/2}}\approx 1.
\]
The last approximation is based on the empirical fact that
for any planet in the solar system the maximum difference  $|\psi(\eta)-1|$ is
$0.026$ (\cite{cook2009interiors} pg 81). Equation (\ref{dr}) with $\psi(\eta)=1$ can be explicitly
integrated
\begin{equation}
  (1+\eta)^{1/2} =\frac{5}{\ov{\rho}\,r^{\,5}}\int_0^r\ov{\rho}(a)\,a^{\,4}\,da.
  \label{dr2}
\end{equation}
From equation (\ref{ovrho2}) we obtain 
\[
\Io=\frac{8\pi}{3}\int_0^\infty r^4\rho(r)dr=\frac{8\pi}{9}\left(\ov\rho(R)R^5-
2\int_0^R r^4\ov\rho(r)dr\right)
\]
and from equation (\ref{cl3})
\[
R\,\epsilon^\prime(R)+2\,\epsilon(R)
=\frac{15\,{\Omega}^{\,2}}{8\pi\,G\, \ov{\rho}(R)}\cdot
\]
From the last three equations we get
\[
\ep(R)=\frac{5}{2}\frac{\Om^2R^3}{Gm}
\Bigg(1+\bigg[\frac{5}{2}\left(1-\frac{3}{2}\frac{\Io}{mR^2}\right)\bigg]^2\Bigg)^{-1},
\]
and,  using equation (\ref{J3}),  we obtain
\begin{equation}
  J_{2v}=\frac{5}{3}\frac{\Om^2R^3}{Gm}\Bigg(
  \bigg\{1+\bigg[\frac{5}{2}\left(1-\frac{3}{2}\frac{\Io}{mR^2}\right)\bigg]^2\bigg\}^{-1}
  -\frac{1}{5}\Bigg)
\label{J4}
\end{equation}  
Finally, using equations (\ref{gmJ}), (\ref{RI}), and (\ref{gmI}) we obtain
the Darwin-Radau approximation for $\gm$:
\begin{equation}
  \frac{\gm}{\gm_\Irm}=\frac{3}{2}\left(\frac{R_\Irm}{R}\right)^5\
  \frac{1+\left[\frac{5}{2}-\frac{3}{2}\left(\frac{R_\Irm}{R}\right)^2\right]^2}
       {4-\left[\frac{5}{2}-\frac{3}{2}\left(\frac{R_\Irm}{R}\right)^2\right]^2}
  \label{gm3}
  \end{equation}
Notice that if $R_\Irm/R=1-\dt$ then the above formula gives 
$\gm/\gm_\Irm=1+\Oc(\dt^3)$, which is equation (\ref{O3}) in the Introduction.

\section{A variational principle for the gravitational modulus
  $\gm$.}
  \label{secvargm}

In this  section,  until otherwise state, 
we  suppose that
\begin{equation}
  \rho\in C^2[0,\infty), \ \rho^\prime(r)<0\ \text{for}\  0<r<R, \ \text{and} \
    \rho(r)=0 \ \text{for}  \ R\le r.
    \label{rhohyp}
\end{equation}    
The gravitational modulus $\gm$ has the following
  integral characterization. 
  Multiply equation (\ref{cl2}) by $\ep(r)$,  integrate over $[0,R]$, and use
  equations (\ref{gmJ}) and  (\ref{J2}) to obtain:
\begin{equation}
 \frac{\gm}{\Io}= -15 G\frac{\int_0^\infty a^5\ov\rho(a)\rho^\prime(a)\ep^2(a)/3da
   +\int_0^\infty\int_{0}^\infty K(a,s)\ep(s)\ep(a) dsda}
      {\left(\int_0^\infty a^5\rho^\prime(a)\ep(a)da\right)^2}
\label{gm1}
\end{equation}
 Moreover, using that 
$\gm_\Irm=\frac{2\Io G}{R_\Irm^5}$ and rescaling the space variables
in the above  integrals  by $R$ , we obtain:
\begin{equation}
  \frac{\gm}{\gm_\Irm}= -\frac{15}{2}\frac{R_\Irm^5}{R^5}
  \frac{\int_0^1 a^5\ov\rho(a)\rho^\prime(a)\ep^2(a)/3da
   +\int_0^1\int_{0}^1 K(a,s)\ep(s)\ep(a) dsda}
      {\left(\int_0^1 a^5\rho^\prime(a)\ep(a)da\right)^2}
\label{gmgmI}
\end{equation}
This expression shows that the nondimensional ratio $\gm/\gm_\Irm$ depends 
on the three variables $m,\Io,$ and $R$ exclusively by means of the
ratio $R_\Irm/R$. In particular, the multiplication of the  density function
of the body by
a constant does not change  the value of $\gm/\gm_\Irm$.

Let $L^2_\rho=L^2([0,R],\rho)$ be the weighted space of  Lesbegue square
integrable functions with inner product given by 
\begin{equation}
  \langle f, g\rangle_\rho=-\int_0^Rf(r)g(r)r^5\rho^\prime(r)dr
\label{innerrho}
\end{equation}
Let $N$ and $M$ be the operators on
$L^2_\rho$ defined by 
\begin{equation}
\begin{split}  
  N[\ep](r)&=
-\left\{\frac{1}{5 r^5} \int_{0}^r a^5\ep(a) \rho^\prime(a)da
+\frac{1}{5}\int_{r}^R \ep(a) \rho^\prime(a)da\right\}\\
&=-\frac{1}{r^5\rho^\prime(r)}\int_{0}^R K(a,r)\ep(a)da\\
   M[\ep](r)&=\frac{\ov\rho(r)}{3}\ep(r)
\end{split}\label{MN}
\end{equation}
Using these operators, Clairaut's equation (\ref{cl2}) can be written as
\begin{equation}
M[\ep](r)-N[\ep](r)=\frac{\Om^2}{8\pi G}\label{cl7}
\end{equation}
Consider the functional $V:L_\rho^2\to \R$ given by
\begin{equation}
  \begin{split}
  V(u)&=-\int_0^R\frac{a^5\ov\rho(a)\rho^\prime(a)}{3}u^2(a)da
  -\int_0^R\int_{0}^R K(a,s)u(s)u(a) dsda\\
  &=\langle u,Mu\rangle_\rho-\langle u,Nu\rangle_\rho\\
  &=   \frac{3}{8\pi G}\int_0^R
   \left[\Phi_2(s)-
     \Phi^\prime_0(s)\frac{2}{3}s u(s)\right]u (s)s^3\rho^\prime(s)ds
  \end{split}
  \label{V1}
   \end{equation}
   where $\Phi_0$ and $\Phi_2$ are the functions defined in equation (\ref{phi1})
   with $u$ replacing $\ep$ in the definition of $\Phi_2$.
   
\begin{lemma} Suppose that $\rho$ satisfies hypothesis {\rm (\ref{rhohyp})}.
  Then, $V(u)>0$ for all $u\in L_\rho^2$, $u\ne 0$.
  \label{Vgt}
\end{lemma}  

\bproof 
The proof of the lemma requires the following.
\begin{proposition}
  There exists a strictly positive function $u_1$ in $L^2_\rho$ and a positive number
  $\lb_1$, they are solutions to the eigenvalue problem $N u_1=\lb_1 M u_1$,
  such that for any $u\in L^2_\rho$, with  $u\ne u_1$, the following inequality holds  
  \[
  \frac{\langle u,N u\rangle_\rho}{\langle u,M u\rangle_\rho}<
  \frac{\langle u_1,N u_1\rangle_\rho}{\langle u_1,M u_1\rangle_\rho}=\lb_1  
  \]
\end{proposition}

\bproof
Consider a new inner product defined by
\[
\langle f,g\rangle_M=\langle f,Mg\rangle_\rho=
-\int_0^Rf(r)g(r)\frac{r^5\rho^\prime(r)\ov \rho(r)}{3}dr
\]
Since  $M[\ep](r)=\frac{\ov\rho(r)}{3}\ep(r)$ and
$\ov\rho(r)\ge \ov\rho(R)>0$ the space $L^2_\rho$ with this new
inner product is also  a Hilbert space that will be denoted as
$L^2_M$.

Note that $M$, defined on $L^2_M$,
is invertible with inverse $M^{-1}u=3u/\ov \rho$.
We define a new operator on $L^2_M$ as $A=M^{-1}N$.
In order to write $A$ more explicitly, 
it is convenient to rewrite the function $K$ given in equation (\ref{K}) as 
\[
K(a,r)=a^5\rho^\prime(a)r^5\rho^\prime(r)P(a,r),
\]
where
\[
P(a,r)=\frac{F(a,r)}{a^5r^5}
 \]
Notice that $P(a,r)=P(r,a)$ and
\[
P(a,r)=\frac{1}{5 r^5}\quad{\rm if}\quad r\ge a, 
\]
which implies that $P$ is continuous for $(a,r)\ne(0,0)$.
Notice that
\[
\begin{split}
A[u](r)&=M^{-1}Nu(r)=-\frac{3}{\ov\rho (r)}\int_0^RP(a,r)u(a)a^5\rho^\prime(a)da
\\&=-\int_0^RQ(a,r)u(a)\frac{\ov\rho (a)}{3}a^5\rho^\prime(a)da
\end{split}
\]
where $Q(a,r)$ is the symmetric positive  function
\[
Q(a,r)=\frac{3}{\ov\rho (r)}\frac{3}{\ov\rho (a)}P(a,r).
\]
This implies that $A$ is symmetric, $\langle u,A v\rangle_M=\langle v,A u\rangle_M$, and
\[
\begin{split}
&\int_0^R\int_0^RQ^2(a,r)
  \frac{\ov\rho (a)}{3}a^5\rho^\prime(a)\frac{\ov\rho (r)}{3}r^5\rho^\prime(r)dadr\\&
  \quad =
\int_0^R\int_0^RP^2(a,r)
a^5\rho^\prime(a)r^5\rho^\prime(r)dadr\\
&\quad=2\int_0^R\int_0^rP^2(a,r)a^5\rho^\prime(a)da\, \, 
r^5\rho^\prime(r)dr\\
&\quad=2\int_0^R\int_0^ra^5\rho^\prime(a)da\,\,
\frac{1}{25 r^5}\rho^\prime(r)dr\le \frac{R^2}{150}\|\rho^\prime\|^2_\infty
\end{split}
\]
Therefore, by theorem VI-23 in \cite{reed1980}, $A$ is a Hilbert-Schmidt operator
in $L^2_M$ which implies that it is bounded and compact. Since $Q(a,R)>0$,
the operator $A$ is
 a strongly positive operator in the sense that it maps a non-negative continuous
 function that  is not identically equal to zero to a strictly positive function.
 All these properties imply that the Krein-Rutman theorem
 (Theorem 7.C in \cite{zeidler1}) can be used to show the existence
 of a unique maximal eigenvalue $\lb_1>0$ associated to a unique positive eigenfunction
 $u_1$ such that $Au_1=\lb_1 u_1$. The maximal eigenvalue of a Hilbert-Schmidt operator
 satisfies the inequalities:
 \[
    \frac{\langle u,A u\rangle_M}{\langle u,u\rangle_M}<
  \frac{\langle u_1,A u_1\rangle_M}{\langle u_1,u_1\rangle_M}=\lb_1,\quad u\ne u_1
  \]
  Then the  proposition follows from:
  $A=M^{-1}N$, $\langle f,g\rangle_M=\langle f,Mg\rangle_\rho$,
  and that a function is in $L^2_M$ if and only if it is in $L^2_\rho$.
  \eproof

  We return to the proof of lemma \ref{Vgt}.  
Let $\ep$ be the 
solution to equation (\ref{cl2}) given in proposition \ref{stat1}.
Multiplying both sides of equation (\ref{cl2}) by $u_1$ and integrating over
$[0,R]$ we  obtain
\begin{eqnarray*}
&&-\frac{\Om^2}{8\pi G}\int_0^\infty u_1(a)a^5 \rho^\prime(a)da=
  \left\{\langle u_1,M\ep\rangle_\rho-
  \langle u_ 1,N\ep\rangle_\rho\right\}\\
  &&=
  \left\{\langle u_1,M\ep\rangle_\rho-
  \langle N u_ 1,\ep\rangle_\rho\right\}=
-(1-\lb_1)\int_0^\infty u_1(a)\frac{a^5\ov\rho(a)\rho^\prime_0(a)}{3}\ep(a)da
\end{eqnarray*}
So, using that $\ep$ and $u_1$ are positive, we obtain
\begin{equation}
1-\lb_1=\frac{3\Om^2}{8\pi G}\frac{\int_0^\infty u_1(a)a^5 \rho^\prime(a)da}
{\int_0^\infty u_1(a)a^5\ov\rho(a)\rho^\prime(a)\ep(a)da}>0.
\label{lb1}
\end{equation}
So,   for any $u\in L_\rho^2$, 
\[
  V(u)=\langle u,M u\rangle_\rho-\langle u,N u\rangle_\rho=
  \lb_1\langle u,M u\rangle_\rho-\langle u,N u\rangle_\rho+
  (1-\lb_1)\langle u,M u\rangle_\rho.
\]
Using that $\lb_1\langle u,M u\rangle-\langle u,N u\rangle_\rho\ge 0$ we get,
for any $ u\in L_\rho^2[0,R]$, 
\begin{equation}
  V(u)\ge
  (1-\lb_1)\int_0^\infty\frac{r^5\ov\rho(r)[-\rho^\prime(r)]}{3}u^2(r)dr>0,
    \label{V2}
  \end{equation}
where $1-\lb_1>0$ is given in equation (\ref{lb1}). This finishes the proof
of the lemma.
\eproof
\vskip.3truecm

The  positivity of the  functional $V$, given in lemma \ref {Vgt}, implies that the
solution to Clairaut's equation (\ref{cl7}) has the following  variational
characterization.
The operator
\begin{equation}
P[u]=(M-N)[u],\quad u\in L^2_\rho\label{P}
\end{equation}
is positive because $V(u)=\langle u,P u\rangle_\rho>0$ if $u\ne 0$.
Clairaut's equation (\ref{cl7}) can be written as
\begin{equation}
P[\ep]=\frac{\Om^2}{8\pi G}
\label{cl9}
\end{equation}
and, up to a multiplicative constant, the solution to this equation is
given by the variational principle (see \cite{stakgold} equation 3.82):
\begin{equation}
  \max_{u\in L_\rho^2} \frac{\langle 1,u\rangle_ \rho^2}{\langle u,P u\rangle_\rho}=
  \max_{u\in L_\rho^2} \frac{\langle 1,u\rangle_\rho^2}{V(u)}=
  \frac{\langle 1,\ep\rangle_\rho^2}{V(\ep)},\label{var1}  
\end{equation}
where the $1$ inside brackets means the constant function equal to one.
This variational characterization      and   equation (\ref{gm1}) 
  imply
\begin{equation}
  \gm= 15G\Io\min_{u\in L_\rho^2}
\frac{V(u)}{\langle 1,u\rangle_\rho^2}= 15G\Io\frac{V(\ep)}{\langle 1,\ep\rangle_\rho^2}
  \label{gm2}
\end{equation}
and, using $\gm_\Irm=(4/5)(Gm/R_\Irm^3)$, 
\begin{equation}
  \frac{\gm}{\gm_\Irm}= \frac{15}{2}R_\Irm^5\min_{u\in L_\rho^2}
  \frac{V(u)}{\langle 1,u\rangle_\rho^2}=
  \frac{15}{2}R_\Irm^5\frac{V(\ep)}{\langle 1,\ep\rangle_\rho^2}
  \label{gm2.5}
\end{equation}

In Appendix \ref{app1} it is shown that\footnote{
  The considerations in the following paragraphs and in Appendix \ref{app1} are
  due to one of the referees of the paper.}
  \begin{equation}
    V(u)=\frac{45}{32\pi^2 G}  U(u)\quad\text{and}\quad
    \langle 1, u\rangle_\rho=-\frac{45}{8\pi\Om^2} U_c(u)
\label{VUc}
  \end{equation}   
  where
  \[
  U(u)=\frac{1}{2}
  \int_{\R^3}\tilde\rho(x)\Phi(x)d x\quad\text{and}\quad
     U_c(u)=
     \int_{\R^3}\tilde\rho(x)\Phi_c(x)d x
   \]
   are the gravitational energy (at second order in the deformation $u$)
   and the centrifugal energy (at first order in $u$), respectively.
   So,  the variational principle in (\ref{gm2}) can be rephrased in terms
   of  the gravitational energy and the centrifugal energy.

   The variational principle (\ref{gm2})
   determines the solution to Clairaut's equation
   up  to a multiplicative constant. There is another variational principle
(see \cite{stakgold} equation 3.80),
   closely related to (\ref{gm2}),  that fully determines the solution to Clairaut's
   integral equation: $\ep$ is a solution to equation (\ref{cl9})
   if and only if it minimizes
   \begin{equation}
     u\to 
     \frac{1}{2}\langle u, P[u]\rangle_\rho-\frac{\Om^2}{8\pi G}\langle 1,u\rangle_\rho
     \label{vp1}
   \end{equation}  
   If this functional is multiplied by $64\pi^2G/45$ and equations (\ref{VUc}) are used
   then this variational principle can be written in a very natural way:
   $\ep$ is a solution to Clairaut's integral equation
      if, and only if, it minimizes the total energy functional 
   \begin{equation}
     u\to E[u]= U[u]+U_c[u] 
     \label{vp2}
   \end{equation}

\subsection{Discontinuous mass distributions with $\rho^\prime(r)\le 0$}

\label{discmass}

The mass distribution of
the ``Preliminary Reference Earth Model'' (PREM) \cite{prem} does not satisfy
hypotheses {\rm (\ref{rhohyp})}.  
From a physical perspective this may not be 
relevant because the density distribution of the  PREM  can be
arbitrarily well approximated
  (pointwise and in the
$L^2$ sense)   by  densities which are smooth and strictly decreasing
in the interval $(0,R)$. In this section we exhibit an approximation scheme that 
allows for  the use of the variational principle in equation (\ref{gm2})
when $\rho:\R\to [0,\infty)$ satisfies the  hypotheses in equation
(\ref{innerrho2}).
%
  It is convenient to  extend $\rho$ to
  $r<0$ as an even function.
  We will regularize $\rho$ using a mollifier.
  Let $f:\R\to \R$ be a    $C^\infty$ positive even  function,
  with support in the interval
  $[-1,1]$, with $\int_\R f(r)dr=1$,  and such that $f^\prime(r)<0$ for $0<r<1$.
  For a given small $\tau>0$,  let $g_\tau:\R\to\R$ be the
  function $g_\tau(r)=\tau(R-|r|)$ for $|r|\le R$ and $g_\tau(r)=0$ for $|r|> R$.
  A regularized density function
  $\rho_\tau$ is defined as
  \[
  \rho_\tau(r)=\frac{1}{\tau}\int_{-\infty}^{\infty}f\big((r-a)/\tau\big)
  \big[\rho(a)+g_\tau(a)\big]da 
  \]
  For any $\tau>0$, the function $\rho_\tau$ is: $C^\infty$, positive,  and even.
  We claim that $\rho_\tau^\prime <0$ if $0<r<R+\tau$. We split the proof of the claim
  into two parts:  $0<\tau\le r$ and $0<r<\tau$.
  For $0<\tau\le r$
  the  derivative of $\rho_\tau$  is
  \[
  \begin{split} \rho_\tau^\prime (r)&=
\frac{1}{\tau^2}\int_{-\infty}^{\infty}f^\prime\big((r-a)/\tau\big)
\big[\rho(a)+g_\tau(a)\big]da\\
&=
\frac{1}{\tau^2}\int_{r-\tau}^{r}f^\prime\big((r-a)/\tau\big)
\big[\rho(a)+g_\tau(a)\big]da\\
&\quad +
\frac{1}{\tau^2}\int_{r}^{r+\tau}f^\prime\big((r-a)/\tau\big)
\big[\rho(a)+g_\tau(a)\big]da.
\end{split}
  \]
  Since: $f^\prime$ is odd,
  $f^\prime\big((r-a)/\tau\big)>0$ ($<0$) for $r<a<r+\tau$ ($r-\tau<a<r$), and
  $\rho(a)+g_\tau(a)$ is positive and  strictly decreasing for $0<a<R$,
  the first term in the
  right hand side of the equation above is negative and it is larger in absolute
  value than the second term, which is positive. Therefore $\rho^\prime_\tau(r)<0$
  for $0<\tau\le r<R+\tau$.
  For $0<r<\tau$ the  derivative of $\rho_\tau$  is
  \[
  \begin{split} \rho_\tau^\prime (r)&=
    \frac{1}{\tau^2}\int_{0}^{2r}f^\prime\big((r-a)/\tau\big)
\big[\rho(a)+g_\tau(a)\big]da\\
&\quad +
\frac{1}{\tau^2}\int_{2r}^{r+\tau}f^\prime\big((r-a)/\tau\big)
\big[\rho(a)+g_\tau(a)\big]da\\
&\quad +
\frac{1}{\tau^2}\int_{r-\tau}^{0}f^\prime\big((r-a)/\tau\big)
\big[\rho(a)+g_\tau(a)\big]da.
\end{split}
  \]
  The integral in the first line is negative by the same argument given in the
  previous case. If we  change the  variables of  the third integral as
  $a\to 2r-a$, then the sum of the second and third integrals can be written as
  \[
\frac{1}{\tau^2}\int_{2r}^{r+\tau}f^\prime\big((r-a)/\tau\big)
\bigg\{\big[\rho(a)+g_\tau(a)\big]-\big[\rho(a-2r)+g_\tau(a-2r)\big]\bigg\}da,
\]
where we used that $f^\prime$ is odd and $\rho+g_\tau$ is even.  This
integral is negative because 
  $f^\prime\big((r-a)/\tau\big)>0$ for $2r<a<r+\tau$ and
  $\rho(a)+g_\tau(a)$ is strictly decreasing for $0<a<R$.
  As $\tau\to 0$ the mollifier $f(r/\tau)/\tau$ tends
  to the Dirac-$\dt$ distribution and therefore $\rho_\tau(r)\to\rho(r)$
   and $\rho^\prime_\tau(r)\to\rho^\prime(r)$
  whenever
  $r$ is a point of continuity of $\rho$. At a  point of discontinuity
$r_j$,    $\rho_\tau(r)\to\chi_j\dt(r-r_j)$ where $\dt$
  is the Dirac $\dt$-measure and $\chi_j$ is the density jump defined in equation
  (\ref{rhod}).

  For a given $\rho$ satisfying hypotheses (\ref{innerrho2}),
  consider the functional $V$ defined on equation (\ref{V1}) 
  restricted to the space of continuous
  functions on $[0,R]$. At a point of discontinuity of $\rho$,
  $\rho^\prime$ must be understood as a $\dt$-distribution (what
  explains the necessity of restricting $V$ to the space of continuous functions).
  Let $V_\tau$ be the same functional defined using the regularized density
  $\rho_\tau$. For a given $u$, standard arguments in the theory of distributions
  show that $\lim_{\tau\to 0}V_\tau(u)\to V(u)$. By lemma \ref{Vgt} $V_\tau (u)>0$
  that implies the following.
  \begin{theorem}
    \label{theor1}
    Suppose that the density distribution $\rho$ satisfies hypotheses
    {\rm  (\ref{innerrho2})} and $u$ is a continuous function on $[0,R]$.
    Then  $V(u)\ge 0$. If $\rho$ has no points of discontinuity then
    the same result is valid for $u\in L^2[0,R]$.
\end{theorem}  
  Notice that if $\rho^\prime(r)=0$ for $r$ in some nonempty open interval in $[0,R]$,
  then $V(u)=0$ for all functions $u$ with support in this interval.

  The same argument shows that equation (\ref{gm2}) is valid under hypotheses
  (\ref{innerrho2}):
  \begin{theorem}
    \label{thvar}
   Suppose that the density distribution $\rho$ satisfies hypotheses
    {\rm  (\ref{innerrho2})} and $u$ is a continuous function on $[0,R]$.
    Then
      \begin{equation}
  \gm\le  15G\Io\frac{V(u)}{\langle 1,u\rangle_\rho^2},
  \label{gm2.7}
      \end{equation}
      where $V$ is given in equation (\ref{V1}) and 
      \[
\langle 1,u\rangle_\rho=-\int_0^Ru(r)r^5\rho^\prime(r)dr.
\]
  If $\ep$ is the solution to the Clairaut's equation then      
    \begin{equation}
 \gm=15G\Io\frac{V(\ep)}{\langle 1,\ep\rangle_\rho^2}
  \label{gm1.5}
    \end{equation}
\end{theorem}
\nd We recall that the solution $\ep$ to the Clairaut's equation is continuous even
  when $\rho$ has points of discontinuity.

\subsection
{Variational principle for piecewise constant mass distributions}
\label{regularization}  
 At first consider the case of a homogeneous body
of constant density $\rho_0$. In this case $-\rho^\prime(r)=\rho_0\dt(r-R)$
and from equations (\ref{innerrho}),  (\ref{V1}), and (\ref{K}) 
\[
\langle 1,u\rangle_\rho=-\int_0^Ru(r)r^5 \rho^\prime(r)dr=
\rho_0\int_0^Ru(r)r^5 \dt(r-R)dr=\rho_0R^5u(R)
\]
and
\[
\begin{split}
  V(u)&=-\int_0^R\frac{a^5\ov\rho(a)\rho^\prime(a)}{3}u^2(a)da
  -\int_0^R\int_{0}^R  \rho^\prime(a)\rho^\prime(r)F(a,r)u(r)u(a) drda\\
  &=\rho_0\int_0^R\frac{a^5\ov\rho(a)}{3}\dt(a-R)u^2(a)da\\&\quad
  -\rho^2_0\int_0^R\dt(r-R)u(r)
  \int_{0}^R  \dt(a-R)F(a,r)u(a) da\, dr\\
  &=\rho_0^2\frac{R^5}{3}u^2(R)-\frac{\rho^2_0}{5}R^5u(R)\int_0^R\dt(r-R)u(r)dr=
    \frac{2}{15}\rho_0^2R^5u^2(R)
\end{split}
\]
So, inequality (\ref{gm2.7}) implies
\[
\gm\le  15G\Io\frac{V(u)}{\langle 1,u\rangle_\rho^2}=
\frac{2\Io G}{R^5}=\frac{4}{5}\frac{Gm}{R^3}
\]
Notice that the right hand side does not depend on $u$ and therefore
the equality holds. This result agrees with that obtained directly from equations
(\ref{cl1}), (\ref{gmJ}),  and (\ref{J2}).

The same computation can be done for a piecewise constant density distribution
as that in equation (\ref{rhoconst}). Using 
$\rho^\prime(r)=-\sum\chi_j\dt(r-r_j)$,
where  from  equation  (\ref{rhod})
\begin{equation}
  \chi_j=\rho_j-\rho_{j-1}\quad\text{for}\quad j=1,\ldots n
  \label{chij}
\end{equation} 
 the result is
\[
\langle 1,u\rangle_\rho=-\int_0^Ru(r)r^5 \rho^\prime(r)dr=
-\sum_{j=1}^n\chi_jr_j^5u(r_j)
\]
and
\[
\begin{split}
  V(u)&=-\int_0^R\frac{a^5\ov\rho(a)\rho^\prime(a)}{3}u^2(a)da
  -\int_0^R\int_{0}^R  \rho^\prime(a)\rho^\prime(r)F(a,r)u(r)u(a) drda\\
  &=-\frac{1}{3}\sum_{j=1}^n\chi_j\ov\rho(r_j)r_j^5u^2(r_j)
-\sum_{j=1}^n\sum_{k=1}^n\chi_j\chi_kF(r_j,r_k)u(r_j)u(r_k)
\end{split}
\]
where, from equation (\ref{K}),
\[
F(r_j,r_k)=  \begin{cases} 
    \frac{r_k^5}{5} & \text{if}\quad r_j\ge r_k \\
    \frac{r_j^5}{5} & \text{if}\quad r_k> r_j 
   \end{cases}
\]
So, for any set of values $\{u(r_1),\ldots,u(r_n)\}\in\R^n$ the following
inequality must hold
\begin{equation}
  \begin{split}
    \gm&\le  15G\Io\frac{V(u)}{\langle 1,u\rangle_\rho^2}\\
  &=
   15G\Io\frac{-\frac{1}{3}\sum_{j=1}^n\chi_j\ov\rho(r_j)r_j^5u^2(r_j)
     -\sum_{j=1}^n\sum_{k=1}^n\chi_j\chi_kF(r_j,r_k)u(r_j)u(r_k)}
   {\left(-\sum_{j=1}^n\chi_jr_j^5u(r_j)\right)^2}
  \end{split}
  \label{varconst}
\end{equation}
with  equality exact at $\{\ep(r_1),\ldots,\ep(r_n)\}$
that is the solution to the discrete Clairaut's equation discussed in Section
\ref{piecewise}.

\section{A Riccati equation associated to  $\gm$ and
inequalities}
\label{riccati}

The Darwin-Radau approximation is obtained from a first order differential,  equation
(\ref{dr}), associated to Clairaut's equation.
In this section, from Clairaut's equation, we derive another
first order differential  equation and from this equation we obtain
a sharp  lower bound for $\gm/\gm_\Irm$.

Equations (\ref{gmI}), (\ref{gmJ}), and (\ref{omep}) can be combined to give  
 \begin{equation}
   \frac{\gm}{\gm_\Irm}=
   \frac{1}{2}\left(\frac{R_\Irm}{R}\right)^5\frac{\zeta}{J_{2v}}, \quad\text{where}
   \quad \zeta=\frac{\Om^2R^3}{Gm}.
    \label{gmJ2}
  \end{equation}
  The substitution of $\ep(R)=-\frac{3}{\ov\rho(R)}y(R)$ and 
$w(R)=-\frac{\Om^2}{8\pi G}$, equations (\ref{cl8}) and (\ref{bc8}), into 
$J_{2v}=\frac{1}{3}(2\ep(R)-\zeta)$, equation (\ref{J3}), gives
  \begin{equation}
\frac{J_{2v}}{\zeta}=\frac{1}{3}\left(\frac{y(R)}{w(R)}-1\right)    
    \label{J52}
  \end{equation}
  The last two  equations give
  \begin{equation}
   \frac{\gm}{\gm_\Irm}=
   \frac{3}{2}\left(\frac{R_\Irm}{R}\right)^5\frac{v(R)}{1-v(R)}\quad\text{where}
   \quad v(r)=\frac{w(r)}{y(r)}.
   \label{v1}
   \end{equation}
   This equation suggests to look for a differential equation for $v$,
   \footnote{Equations (\ref{gmk2}) and  (\ref{v1}) imply that  $v(R)$ is
    entirely determined by 
the static fluid Love number $k_2$
\begin{equation}
  k_2=\frac{1-v(R)}{v(R)}
  \label{vk2}
\end{equation}} which is
   readily obtained from equations (\ref{cl8}) and (\ref{bc8}):
   \begin{equation}
     v^\prime=\frac{5}{r}v(1-v)+\frac{3}{5}\frac{\rho^\prime}{\ov\rho}\quad\text{with}
     \quad v(0)=1
     \label{v2}
   \end{equation}  
   Notice that $v$ satisfies a Riccati equation
   (or a non homogeneous logistic equation).
   
  If $\rho$ has a point of discontinuity at $r_j$ then the jump  condition
   is
\begin{equation}
     \Delta v(r_j)=\frac{3}{5}\frac{\chi_j}{\ov\rho(r_j)},\quad 
     \chi_j=\rho(r_j)-\lim_{r\to (r_{j})_-} \rho(r)< 0
     \label{jumpv}
   \end{equation}
   
   Proposition \ref{stat1} states  that $\ep(r)$  does not change sign and so
   $y(r)=-\frac{\ov\rho(r)}{3}\ep(r)$.
   In the proof of Proposition (\ref{stat1})  we
   showed that $w(r)$ does not change sign either (see equation (\ref{wpos})) and, 
   since $v(0)=1$, 
   $v=w/y$ is always positive. Since  $\rho^\prime/\ov\rho\le 0$ and
   $v(1-v)$ is positive for $0<v<1$ and negative for $v>1$, the solution
   $v(r)$ of equation (\ref{v2}) satisfies
   \begin{equation}
     0<v(r)\le 1,\quad 0\le r\le R.
     \label{vin1}
   \end{equation}
   \begin{theorem}
    \label{thlow}  
 If $\rho$ satisfies the hypothesis in equation {\rm(\ref{innerrho2})} then,  for given values of $m$,  $\Io$,  and $R$, 
\begin{equation}
  \frac{\gm}{\gm_\Irm}\ge
  \frac{3}{2}\, \left(\frac{R_\Irm}{R}\right)^5\,
  \left[\frac{5}{3}\left(\frac{R}{R_\Irm}\right)^2-1\right],
\label{gmlow}     
\end{equation} 
where the equality is verified for a ``thick shell Roche model'' described in Section
{\rm \ref{roche}} equation {\rm (\ref{limgm})},
which do not satisfy hypothesis {\rm(\ref{innerrho2})}.
So,  the value of
$\gm/\gm_\Irm$ is minimum for 
the Roche model
that   consists of a 
point at the origin with mass $m_0=[1-(R_\Irm/R)^2]m$ and a surrounding
homogeneous layer with  mass $m_1=(R_\Irm/R)^2m$.
\end{theorem}
\bproof
In the following we will assume that
$\rho$ is $C^2$.
The same regularization argument presented in Section \ref{regularization}
implies that the theorem holds for discontinuous densities
as those in  hypothesis
(\ref{innerrho2}).

Equation (\ref{v1}) implies that inequality (\ref{gmlow}) holds if, and only
if,
\begin{equation}
  v(R)\ge 1- \frac{3}{5}\left(\frac{R_\Irm}{R}\right)^2
  \label{vrin}
\end{equation}
In order to estimate $v(R)$, we multiply equation (\ref{v2}) by $r^5\ov \rho(r)$ and
integrate its left hand side by parts
\[
  \begin{split}
    \int_0^R\ov \rho(r)r^5 v^\prime(r) dr&=  \ov \rho(R)R^5 v(R) -
    \int_0^R\Big[\big(\ov \rho(r)r^5\big)^\prime+
    \big(\ov \rho(r)r^5\big)^\prime\big(v(r)-1\big)\Big]dr\\
   &=   \ov \rho(R)R^5 v(R) -\ov \rho(R)R^5
    -\int_0^R
    \big(\ov \rho(r)r^5\big)^\prime\big(v(r)-1\big)dr
  \end{split}
\]  
Then, using that
\[
  \frac{3}{8\pi} \Io=\int_0^R r^4\rho(r)dr=-\frac{1}{5}
  \int_0^R r^5\rho^\prime(r)dr\quad\text{and}\quad
  \frac{9}{8\pi}\frac{\Io}{\ov\rho(R) R^5}=
  \frac{3}{5}\left(\frac{R_\Irm}{R}\right)^2
\]
we obtain
\[
  v(R)=1-\frac{3}{5}\left(\frac{R_\Irm}{R}\right)^2
  +\frac{1}{R^5\ov\rho(R)}
\int_0^R \big(1-v\big)
   \big[5r^4\ov \rho v- (r^5\ov \rho)^\prime\big]dr
  \]
So inequality (\ref{vrin}) holds if
  \[
    5r^4\ov \rho v- (r^5\ov \rho)^\prime=r^4(5\ov \rho v -3\rho-2\ov\rho)=r^4H(r)\ge 0,
 \]   
 where we used $\ov\rho^\prime(r)=\frac{3}{r}\bigl[\rho(r)-\ov\rho(r)\bigr]$,
 equation (\ref{ovrho2}).
 We will show that $H(r)\ge 0$ for $0\le r\le R$.
 
 Notice that
 \[
 H(0)= 5\ov \rho(0) v(0) -3\rho(0)-2\ov\rho(0)=0,
\]
and, after some computation using equation (\ref{v2}), the definition of $H$,
and equation (\ref{ovrho2})
 \begin{equation}
  rH^\prime(r)=-5v(r)H(r)+6\big[\ov\rho(r)-\rho(r)\big].
  \label{rH}
\end{equation}  
Let   $\tilde r=\sup_{r\ge 0}\big\{\big[\ov\rho(r)-\rho(r)\big]=0\big\}$.
Notice that   $\ov\rho(r)-\rho(r)>0$ for $r>\tilde r$.
The differential equation for $H$, its differentiability at $r=0$,
and the initial condition $H(0)=0$  implies that $H(r)=0$ for $0\le r\le \tilde r$.
For the same reasons $v(r)=1$ for $0\le r\le \tilde r$, see equation (\ref{v2}).

We will show  that there exists $\ov r>\tilde r$
sufficiently close to $\tilde r$ such that $H(\ov r)>0$ and 
$H(r)\ge 0$ for $\tilde r<r\le \ov r$.
Since $v(\tilde r)=1$,  there exist $\hat r>\tilde r$ sufficiently close to
$\tilde r$ such that, for $\tilde r<r<\hat r$,
$1-5v(r)<0$. Now, suppose that there exists a value of $r_*\in (\tilde r,\hat r)$
such that $H(r_*)< 0$. Then there exist $r_{**}\in [\tilde r,r_{*})$
(possibly $r_{**}=\tilde r$)
such that $H(r)<0$ for $r_{**}<r\le r_*$ and $H(r_{**})=0$.
The integration of  equation (\ref{rH}) gives
\[
  r_*H(r_*)=\int_{ r_{**}}^{r_*}\big(1-5v(a)\big)H(a)da+
  6\int_{\tilde r}^r\big[\ov\rho(a)-\rho(a)\big]da,
\]
which is impossible because the left hand side of this equation is strictly negative
and the right hand side is strictly positive.
So $H(r)\ge 0$ for $r\in (\tilde r, \hat r)$
and again integration of equation (\ref{rH}) gives that $H(\ov r)>0$ for
some $\ov r>\tilde r$ sufficiently close to $\tilde r$.
Now we  claim that $H(r)>0$ for $r\ge \ov r$. Indeed, if there exists
$r>\ov r$ such that $H(r)=0$,  then equation (\ref{rH}) implies  
$rH^\prime(r)=6(\ov\rho(r)-\rho(r))> 0$ which is impossible. Therefore
$H(r)\ge 0$ for $0\le r\le R$.
\eproof

The next theorem establishes  
upper bounds for $\gm/\gm_\Irm$. Let $\ov\Gm:(0,1]:\to \R$ be the function 
\[
  \ov\Gm(R_\Irm/R)=\sup_{\rho}\{\gm/ \gm_\Irm: R_\Irm/R\  \text{is fixed}\},\
\]
where the supremum is taken over all $\rho's$ that
satisfies hypothesis (\ref{innerrho2}).
This  function  is non increasing due to the ill defined concept of geometric
radius
for bodies with  low density external shells (see \cite{baschek1991parameters}
for a discussion on the definition of the radius of a star).
The proof that $\ov\Gm$ is non increasing is based on the following argument:
a density  distribution that is positive only for $r<\hat R$
can be extended to a larger radius $R$ adding a negligible layer of mass that 
does not change the value of $\gm/\gm_\Irm$ but decreases the value of
$R_\Irm/R$. A detailed proof is the following.
Suppose that for a certain value of $\hat R_\Irm/\hat R$ there exists a density
$\hat \rho$ for which $\hat \gm/\hat\gm_\Irm=\ov\Gm(\hat R_\Irm/\hat R)$
(if the supremum of $\gm/\gm_\Irm$
is not realized by any density $\hat \rho$ then $\hat \rho$ must be substituted
for a maximizing sequence). For a  small value of $\tau>0$ let
$R_\tau$ be the largest value of $r$ such that $\hat \rho(r)> \tau$ for
$r< R_\tau$. For $R>\hat R$ consider the  new density function
given by: $\rho_\tau (r)=\hat \rho (r)$ for $r< R_\tau$, 
$\rho_\tau(r)=\tau$ for $ R_\tau\le r< R$, and $\rho_\tau(r)=0$ for $r\ge R$.
Notice that  $\rho_\tau\to \hat \rho$ as $\tau\to 0$ uniformly in the interval
$[0,R]$. Therefore all the quantities $m_\tau$,
${\rm I}_{\circ\tau}$, $R_{\Irm\tau}$,
$\gm_{\Irm \tau}$ and $\gm_\tau$ tends to those respective quantities
of $\hat \rho$  as $\tau\to 0$ (note that for $r>\hat R$
the solution $\hat v$ to equation (\ref{v2}) satisfies 
$\frac{1}{r^5}\frac{\hat v(r)}{1-\hat v(r)}=$constant such that 
$\frac{\gm}{\gm_\Irm}$ in equation (\ref{v1}) remains constant).
Therefore $\ov\Gm(\hat R_\Irm/R)\ge \ov\Gm(\hat R_\Irm/\hat R)$ for $R\ge \hat R$
which implies that $\ov\Gm$ is non increasing. Since for a homogeneous
body $R=R_\Irm$ and $\gm/\gm_\Irm=1 =\ov\Gm(1)$,
we obtain that $\ov\Gm(R_\Irm/R)\ge 1$ for $R_\Irm/R\le 1$. A body
that realizes $\gm/ \gm_\Irm=1$ for any $R_\Irm/R\in (0,1]$ is the
homogeneous core Roche model of Section \ref{roche}.
 The next theorem
shows that the upper bound $\gm/\gm_\Irm\le 1$ is almost correct.

   \begin{theorem}
     \label{thup}
     If $\rho$ satisfies the hypothesis in equation {\rm(\ref{innerrho2})}
     then, for given values of $m$ and   $\Io$,   
\[     
\frac{\gm}{\gm_\Irm}\le
\sqrt{\frac{35}{39}}\,\frac{8575}{8112}\approx 1.001401\quad
\text{for all}\quad \frac{R_\Irm}{R}\in (0,1]
\]
\end{theorem}
\bproof
For a given   $\rho$, 
consider a family of auxiliary density functions  $\rho_s$, $s>0$,
 obtained from $\rho$ by means of a cutoff  at radius $s$, namely
$\rho_s(r)=\rho(r)$ for $r<s$ and $\rho_s(r)=0$ for $r\ge s$. The idea of the proof
relies upon the study of the function $\Gm(s)=\frac{\gm}{\gm_\Irm}(s)$  for
which a differential equation will be written.

The solution $v_s(r)$ to equation (\ref{v2}), for the density $\rho_s$,
coincides with the solution $v(r)$ to the same equation, for the density $\rho$,
as far as $r<s$. At $r=s$
a  jump, possibly null, must be added to $v_s$ and according to
equation (\ref{jumpv}) 
\begin{equation}
v_s(s)=v(s_-)-\frac{3}{5}\frac{\rho(s_-)}{\ov\rho(s)},
\label{vs1}
\end{equation}
where $s_-=\lim_{\{r\to s,\,  r<s\}}r$.
For $r>s$, $\rho_s^\prime=0$ and equation (\ref{v2}) can be explicitly solved
\[
 v_s(r)= \frac{v_s(s) r^5}{
   s^5\big[1-v_s(s)\big]+v_s(s)r^5}
\]

It is a remarkable fact that   
$\rho/\ov\rho$ satisfies a differential equation
which is similar to equation (\ref{v2}).
Indeed,   equation (\ref{ovrho2})  implies 
    \begin{equation}
  \left(\frac{\rho}{\ov\rho}\right)^\prime=
  \frac{3}{r}\frac{\rho}{\ov\rho}\left[1-\frac{\rho}{\ov\rho}\right]+
  \frac{\rho^\prime}{\ov\rho}
\label{ovrho3}
 \end{equation}
     and from this follows a very symmetric form of equation  (\ref{v2}):
\begin{equation}
     v^\prime=\frac{5}{r}v(1-v)+\frac{3}{5}\left(
 f^\prime-\frac{3}{r}f(1-f)\right)\quad\text{where}\quad
f=\frac{\rho}{\ov\rho}
     \label{v3}
   \end{equation}
   or
   \begin{equation}
     \frac{d}{dr}\left(v-\frac{3}{5}\frac{\rho}{\ov\rho}\right) =
     \frac{5}{r}v(1-v)-\frac{9}{5r}f(1-f)
     \label{v4}
   \end{equation}  
  Notice that the left hand side of this last equation is the derivative
  of the function $v_s(r)$ at $r=s$ as given in equation (\ref{vs1}).
  So, we define a new variable
  \[
    z(r)=v(r)-\frac{3}{5}\frac{\rho(r)}{\ov\rho(r)}=
    v(r)-\frac{3}{5}f(r)
  \]
  and from equation (\ref{v4}) and the initial  values
$v(0)=1$ and  $f(0)=\rho(0)/\ov\rho(0)=1$
  we obtain
  \begin{equation}
    z^\prime=\frac{5}{r}z(1-z)+\frac{6}{5}\frac{f}{r}(1-5z), \quad z(0)=\frac{2}{5}. 
    \label{z1}
    \end{equation}
  We remark that $v_s(s)=z(s)$. Let $\bt(s)$ be the the inertial radius
  of $\rho_s$ divided by $s$, namely
  \[
    \bt^2(s)=\frac{5}{3}\frac{\int_0^sa^4\rho da}{s^2\int_0^sa^2\rho da}
  \]
  A computation using the definitions $f=\rho/\ov \rho$ and
  $\ov\rho=\frac{3}{r^3}\int_0^ra^2\rho da$ gives
  \begin{equation}
    \frac{d}{ds}\bt^2=\frac{1}{s}\big(-(3f+2)\bt^2+5f\big)
    \label{bt2}
  \end{equation}  
  All these definitions and equation (\ref{v1}),
namely $   \frac{\gm}{\gm_\Irm}=
   \frac{3}{2}\left(\frac{R_\Irm}{R}\right)^5\frac{v(R)}{1-v(R)}$,
  imply
  that $\Gm(s)$, which is the value
  of $\frac{\gm}{\gm_\Irm}$ for the density function $\rho_s$,  
is given by 
  \begin{equation}
   \Gm(s)=
   \frac{3}{2}\bt^5(s)\frac{z(s)}{1-z(s)}.
   \label{v5}
 \end{equation}
Notice that $\bt(0)=1$ and $z(0)=2/5$ imply $\Gm(0)=1$. 
Differentiating both sides of  equation (\ref{v5}) with respect to $s$ 
and using equations (\ref{z1}) and (\ref{bt2})
   we obtain
   \begin{equation}
     \frac{d}{ds}\Gm=\frac{f}{10 s \bt^5}
     \Big(18\bt^{10}+(125\bt^3-111\bt^5)\Gm-32\Gm^2\Big), \quad \Gm(0)=1,\label{Gm1}
   \end{equation}  
   which is the desired equation. We recall that
   for any value of $s\ge 0$ the following
   inequalities are verified:
   \begin{equation}
     0\le f(s)\le 1\quad \text{and}\quad  0<\bt(s)\le 1
     \label{fgin}
   \end{equation}
   with $f(s)=0$, if and only, if $\rho(s)=0$.

   \begin{wrapfigure}[12]{r}{0.35\textwidth}
  \begin{center}
    \includegraphics[width=0.35\textwidth]{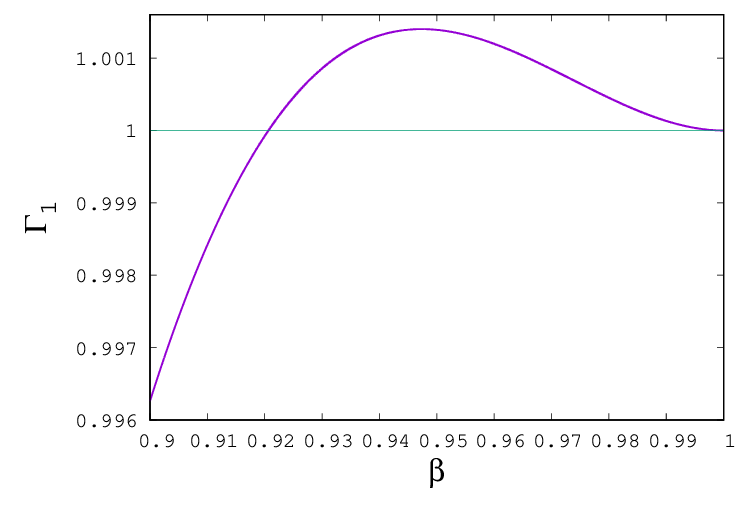}
  \caption{$\Gm_1(\bt)$.}
  \label{fup}
\end{center}

\end{wrapfigure}
   The right hand side of equation (\ref{Gm1}) has two factors:
   $\frac{f}{10 s \bt^5}$, which is greater or equal to zero  for $s>0$, and
    $P(\bt,\Gm)=18\bt^{10}+(125\bt^3-111\bt^5)\Gm-32\Gm^2$, which is
    a quadratic polynomial in $\Gm$.
    For a given value of $\bt\in(0,1]$, $\Gm\to P(\bt,\Gm)$ has a  positive root
    $\Gm=\Gm_1(\bt)$ given by 
{\footnotesize  \[
\Gm_1(\bt)=\bt^3\frac{125-111\bt^2+5 \sqrt{5} \sqrt{117\bt^4-222\bt^2+125}}{64},
\]}

\nd
and another negative root such that $P(\bt,\Gm)<0$ for $\Gm>\Gm_1(\bt)$
and $P(\bt,\Gm)>0$ for $0<\Gm<\Gm_ 1(\bt)$. The graph of $ \Gm_1(\bt)$ 
for $\bt\in(0.9,1)$ is given in Figure \ref{fup}.
 The  point of maximum $\tilde \bt$
 of $\Gm_1(\bt)$ can be  calculated in the following way.
 Equations 
$\frac{d}{d\bt}P(\bt,\Gm_1(\bt))=0$
and  $\frac{d}{d\bt}\Gm_1(\bt)=0$  give
\[
\Gm_1(\bt)=\frac{180\bt^9}{-375\bt^2+555\bt^4}
\]
which can be  substituted into   $P(\bt,\Gm_1(\bt))=0$ to give an
equation for $\bt$
\[
P(\bt,\Gm_1(\bt))=  -168750 (\bt-1) \bt^{14} (\bt+1) \left(39 \bt^2-35\right)=0
\]  
The only root to this equation in the interval $(0,1)$ is
$\tilde \bt=\sqrt{35/39}$. The value of
$\Gm_1(\bt)$ at $\bt=\tilde \bt$ is
$\frac{8575}{8112}\sqrt{\frac{35}{39}}$.

The above argument shows that  for:
$\Gm=\frac{8575}{8112}\sqrt{\frac{35}{39}}$, $0<s$, $0<\bt\le 1$,
and $0\le f\le 1$,
the right hand side of equation
(\ref{Gm1}), and therefore $\frac{d}{ds}\Gm$,  is less than  or equal to zero.
So no solution to equation (\ref{Gm1}) that starts at $\Gm(0)=1$, for any admissible
$f$ and $\bt$ that satisfy inequalities (\ref{fgin}),
can cross above the line 
$\Gm=\frac{8575}{8112}\sqrt{\frac{35}{39}}$,  which ends the proof of the theorem. 
\eproof

\nd {\it Remark:}  There are density
functions  for which $\gm/\gm_\Irm$ is larger than one: 
a polytrope of index $n=0.4604$ has $R_\Irm/R=0.9102$ and
$\gm/ \gm_\Irm\approx 1.0003> 1$ (numerically estimated); and  the 
parabolic density function $\rho(r)=1-r^2$,
for $0\le r\le \sqrt{28/48}$, and $\rho(r)=0$,
for $r\ge \sqrt{28/48}$, has $R_\Irm/R=0.947331$ and
$\gm/\gm_\Irm \approx 1.0008$ (numerically estimated).
I believe that the upper bound
$\frac{\gm}{ \gm_\Irm}\approx 1.001401$ is sharp, namely
there are density functions for which the value of $\gm/\gm_\Irm$ gets arbitrarily close
to $1.001401\ldots$. If this is true, then it is an interesting
mathematical problem to determine the limit density profile that maximizes
$\gm/\gm_\Irm$. Both, the fact that the upper bound  of $\gm/\gm_\Irm$
as a function of $R_\Irm/R$ is non increasing and that
$\gm/\gm_\Irm=1$ for $R_\Irm/R=1$,  shows that our upper bound as a function
of $R_\Irm/R$ 
can be   sharp only for $R_\Irm/R\le \sqrt{35/39}$.

\begin{figure}[h]
\centering
\begin{minipage}{0.5\textwidth}
\centering
\includegraphics[width=\textwidth]{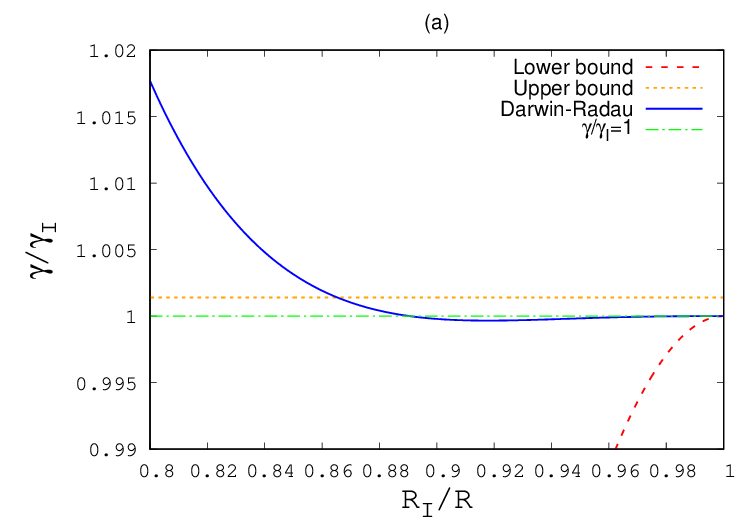}
\end{minipage}\hfill
\begin{minipage}{0.5\textwidth}
\centering
\includegraphics[width=\textwidth]{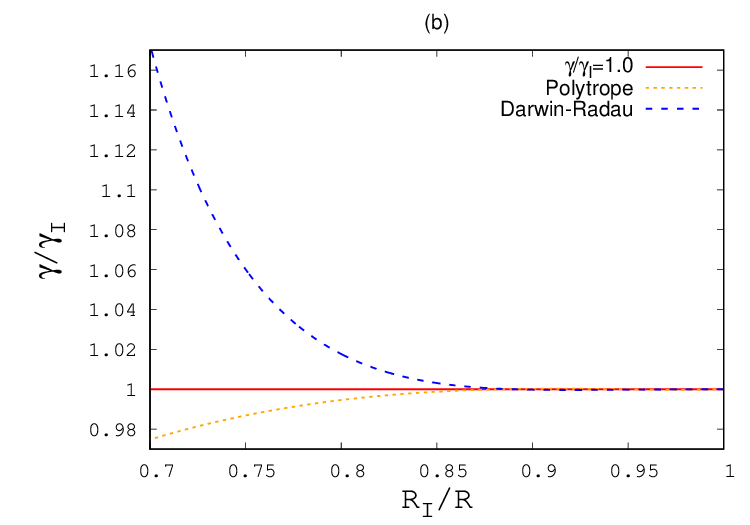}
\end{minipage}\hfill
\caption{(a) Comparison between the lower bound in Theorem
  \ref{thlow}, the upper bound in  Theorem
  \ref{thup}, and the Darwin-Radau approximation given in equation
  (\ref{gm3}). The Darwin-Radau approximation violates the upper bound
  if  $\frac{R_\Irm}{R}<0.86534\ldots$.
  (b)  Comparison between the numerically computed value of
  $\gm/\gm_\Irm$ for a polytrope and the same quantity computed using the Darwin-Radau
  approximation. Notice that for $0.7\le R_\Irm/R<1$ the value 
  $\gm/\gm_\Irm=1$ approximates better the actual value of this quantity
  for the polytrope than the Darwin-Radau approximation.} 
\label{fradau}
\end{figure}

\begin{figure}[t]
\begin{center}
\includegraphics[width=0.9\textwidth]{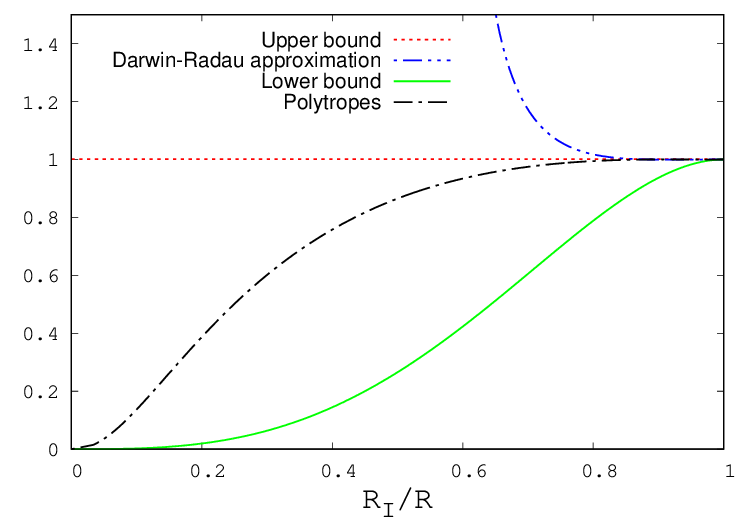}
\end{center}
\caption{Four graphs of the normalized gravitational  modulus $\gm/\gm_\Irm$
  as a function of the normalized inertial radius $R_\Irm/R$: 
  lower and upper bounds given in Theorems (\ref{thlow}) and (\ref{thup}),
  respectively, Darwin-Radau
  approximation in equation (\ref{gm3}),
  and polytropes (see section \ref{polysec}).} 
\label{figin1}
\end{figure}

Theorem \ref{thup} establishes a limit for the validity of the Darwin-Radau theory.
Indeed,  as illustrated in Figure \ref{fradau} (a),
the value of $\gm/\gm_\Irm$  in equation (\ref{gm3}) given by
the Darwin-Radau approximation is smaller than that in the upper bound
given in theorem \ref{thup} if, and only if, 
\begin{equation}
  \frac{R_\Irm}{R}>0.86534\ldots,\quad \text{(Validity
    of the Darwin-Radau approximation).}
  \label{limaradau}
\end{equation}
From Figure \ref{fradau} (b) it is possible to see that for a polytrope, within
the range
$0.7\le R_\Irm/R<1$,
the approximation $\gm=\gm_\Irm$ has a maximum
relative error of the order of  $3\%$ while the Darwin-Radau approximation
has a relative error of the order of $20\%$.
  Figure \ref{figin1} shows  $\gm/\gm_\Irm$ as a function
  of $R_\Irm/R\in(0,1)$ for:   the Clairaut's approximation, for the polytropes,
  for the  thick shell
  Roche models presented
  in Section \ref{roche}, which are the lower bounds for
  $\gm/\gm_\Irm$, and for the upper bound given in Theorem \ref{thup}.

\section{Computation of $\gm$ for some celestial bodies}

\label{computation}

In this section we compute the values of $\gm$ for some  bodies in the solar system
solving numerically  Clairaut's equation according to the algorithm
described in the paragraph above equation (\ref{trans2}).
In the literature there is more than
one  proposal of mass-distribution
for  the same body, in most cases  we just chose one.  Our  
goal is to compare the values obtained with: the direct integration
of the Clairaut's equation,  the Darwin-Radau approximation, the
upper and lower bounds in Theorems \ref{thlow} and \ref{thup}, and the
value $\gm_{ob}=\frac{\Io}{mR^2}\frac{\Om^2}{J_{2v}}$ (equation (\ref{gmJ})),
where $\Om$ and $J_{2v}$ are values found in the literature, which were estimated from
observations.

In principle,  the  values of $\gm_{ob}$ do not have  to
match  the   value computed using Clairaut's equation for three reasons.
The first is that in Clairaut's theory only gravitational
forces are taking into account while $\gm_{ob}$ 
is due to gravity plus solid and fluid  elastic forces. The dominance of
the gravitational forces over the elastic forces tends to increase as
the body increase. The second reason is that tide-dissipation is slowing
down the spin of  celestial bodies, so viscous forces,  both in the fluid
and in the solid part,  may offset the system from equilibrium
(this was the explanation found in \cite{mckenzie1966} for the difference
between the Earth flattening predicted under the hydrostatic hypothesis
and the observed one). The third reason is that Clairaut's theory is
of first order in the small parameter $\frac{\Om^2R^3}{Gm}$, so, as this parameter
increases, higher order corrections become more important. In spite of all
these remarks,  the values of $\gm_{ob}$ are reasonably well
approximated by the values found using Clairaut's theory.

\begin{figure}[hptb]
\centering
\begin{minipage}{0.48\textwidth}
\centering
\includegraphics[width=\textwidth]{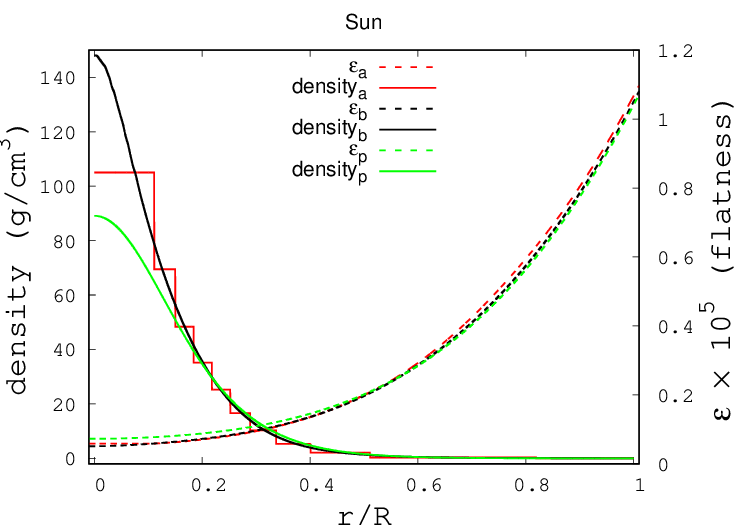}
\includegraphics[width=\textwidth]{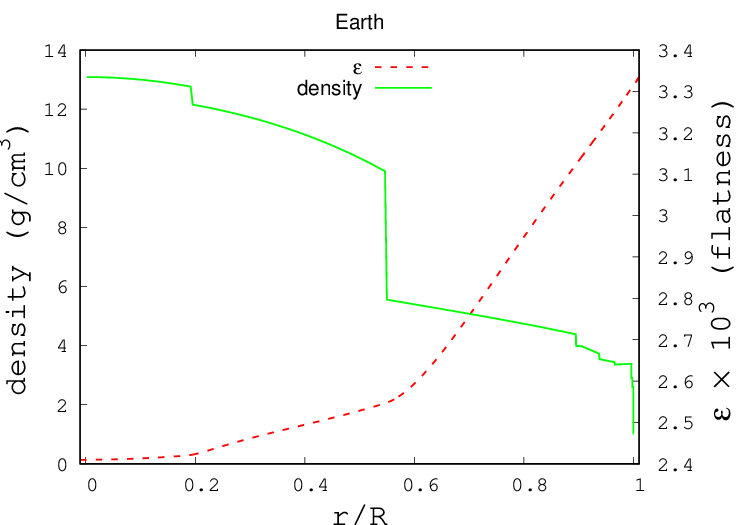}
\includegraphics[width=\textwidth]{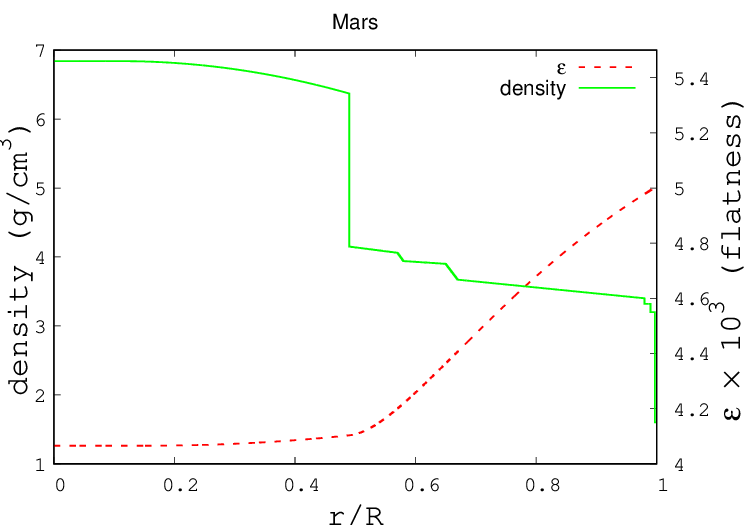}
\end{minipage}\hfill
\begin{minipage}{0.48\textwidth}
\centering
\includegraphics[width=\textwidth]{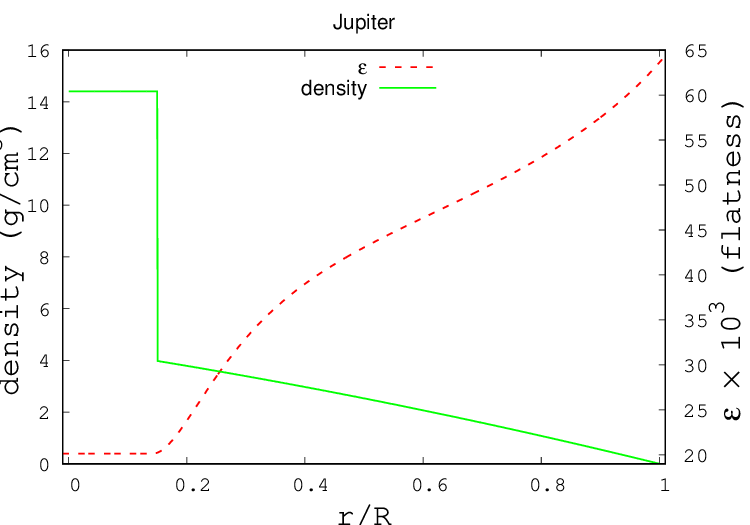}
\includegraphics[width=\textwidth]{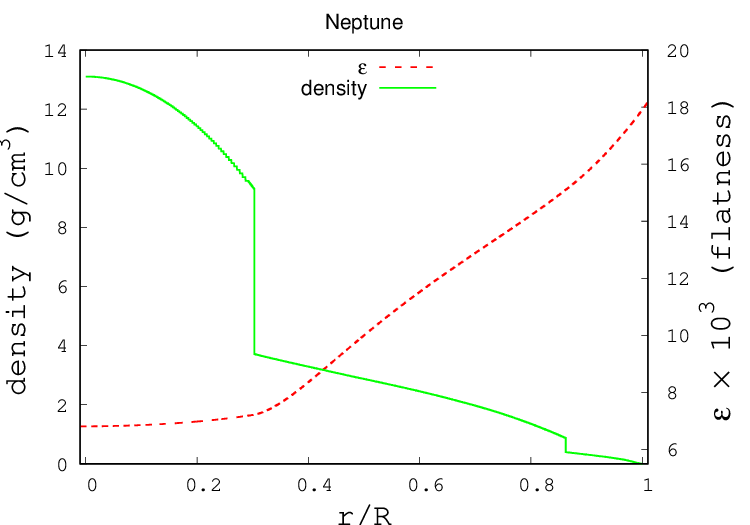}
\includegraphics[width=\textwidth]{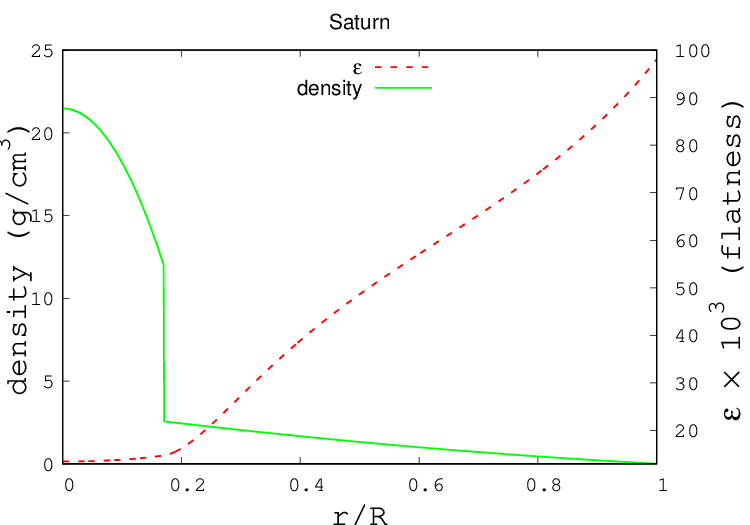}
\end{minipage}
\caption{Mass density distribution $\rho$ and flatness $\ep$,  which
  was obtained from the
  numerical integration of Clairaut's equation, for six of the bodies
  in Table \ref{tab2}. Three different density models were used for the Sun:
  the index ``a'' refers to   the density distribution in \cite{burvsa}
  (remark (j)), the index 
  ``b'' to   the density distribution in \cite{bahcall1988}
  (remark (k)), and the index ``p'' to the density of a polytrope
  of index  $n=3.060$, which corresponds to $R_\Irm/R=0.423$, and mass
  $m=1.9885\times 10^{30}$kg  (remark (k)).
 The density
  distribution for the Earth, Mars,  Jupiter,   Saturn, and Neptune were taken, respectively,
  from \cite{prem}, \cite{zharkov2005} (model M13),
  \cite{hubbard2016}, \cite{helled2017internal},
  and  \cite{nettel2013uran}
  (model N1).} 
\label{dep}
\end{figure}

\begin{figure}[hptb]
\begin{center}
\includegraphics[width=\textwidth]{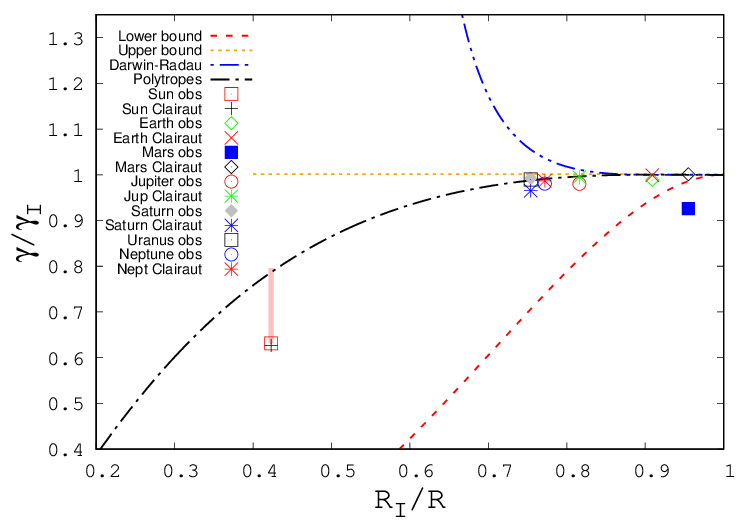}
\end{center}
\caption{The four graphs in this figure are:
  the lower and upper bounds for $\gm/\gm_\Irm$, the Darwin-Radau approximation,
  and $\gm_{P}/\gm_\Irm$ for polytropes. The points indicated with ``obs'' represent
  the values of $(R_\Irm/R, \gm_{ob}/\gm_\Irm)$, where $R_\Irm/R$ and $\gm_{ob}/\gm_\Irm$
  are given in Tables \ref{tab1} and   \ref{tab2}, respectively, and those
  points indicated with ``Clairaut'' represent $(R_\Irm/R, \gm_{C}/\gm_\Irm)$,
  where $\gm_{C}/\gm_\Irm$ is given in Table \ref{tab2}. The vertical line
  represents 
  the possible values of $\gm_{ob}/\gm_\Irm$ for the Sun as 
  the value of  $J_2$ varies
  from $J_2=0.182\times 10^{-6}$ \cite{inpop08} to
  $J_2=0.2295\times 10^{-6}$\cite{inpop17a}, this last value being that
  used to obtain $\gm_{ob}/\gm_\Irm=0.6314$ (see remark (l)).
  For an enlargement of the region $0.7\le R_\Irm/R\le 1$ see Figure
\ref{ampl}. }
\label{sum}
\end{figure}

\begin{figure}[hptb]
\begin{center}
\includegraphics[width=\textwidth]{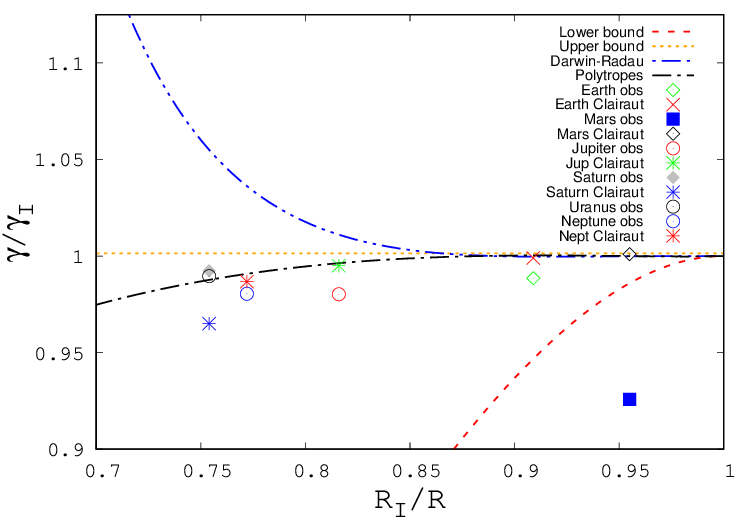}
\end{center}
\caption{ Magnification of the region  $0.7\le R_\Irm/R\le 1$ of  Figure
\ref{sum}.}
\label{ampl}
\end{figure}

The results are summarized in:  Table \ref{tab1}, which contains
the data used in the computations;  Tables \ref{tab2}, which contain the values
of $\gm$ obtained in different ways; in Figure \ref{dep}; which contains
the density functions and the value of the flatness for six
 of the bodies in Table \ref{tab2};
and Figures \ref{sum} and \ref{ampl} that summarize all the data in this section.

{\it Remarks and  notation:}
\begin{itemize}
\item[a)] The gravitational constant is
$G=6.67408\times 10^{-11}$ m$^3$ kg$^{-1}$ s$^{-2}$   
\item[b)]  $\gm_{C}$ is the value computed integrating numerically Clairaut's equation.
\item[c)]
  $\gm_\Irm=(4/5)(Gm/R_\Irm^3)$ is the value of $\gm$ for a homogeneous body
  with mass $m$ and  radius equal to the radius of inertia
  $R_\Irm=\sqrt{\frac{5\Io}{2m}}$.
\item[d)]  $\gm_{DR}$ is the value of $\gm$ computed using the equation of 
   Darwin-Radau, which  requires $R_\Irm/R>1/\sqrt{3}$ (at
  $R_\Irm/R=1/\sqrt{3}$ the denominator of the right-hand side of equation
(\ref{gm3}) becomes zero), so it cannot be computed for  the Sun.
\item[e)]  $\gm_{P}$ is the value of $\gm$ under the hypothesis that
  the  body  is made of a polytropic
  fluid with an index determined by the ratio $R_\Irm/R$ (see Section
  \ref{polysec}). 
\item[f)]
  $\gm_{ob}=\frac{\Io}{mR^2}\frac{\Om^2}{J_{2v}}=\frac{\Io}{mR_e^2}\frac{\Om^2}{J_{2}}$
  is the value of $\gm$
where $\Om$ and $J_{2}$ are numbers found in the literature, which were estimated from
observations.  All
the equatorial radii $R_e$ used in this paper were taken from
\cite{yoder1995astrometric}.
As  remarked above  $\gm_{ob}$ may  not  represent
the same physical quantity as
  $\gm_{C}$, $\gm_P$,  or $\gm_{DR}$.

\item[g)] The value  $C/(mR^2)$, where $C$ is the polar moment
  of inertia, is more frequently found  in the literature than  $\Io/(mR^2)$.
  The two quantities are related by
  \[
    \frac{\Io}{mR^2}=\frac{C}{mR^2}-\frac{2 }{3}J_{2v}=
    \frac{C}{mR^2}-\frac{2}{3}\frac{R_e^2}{R^2}J_2
    \]

\item[h)] For Mars,  the value of $\gm_{ob}/\gm_\Irm=0.9258$ is considerably
  smaller
  than that obtained from the Darwin-Radau approximation $\gm_{DR}/\gm_\Irm=0.9999$.
  This difference is discussed in  \cite{zharkov2005} (see p. 368) and it may be
  caused by non gravitational internal  tensions.  

\item[i)] {\it The Sun I.} It was more difficult to obtain
  the several values of $\gm$ for the Sun than for the other bodies.
  The Sun has a ratio $R_\Irm/R\approx 0.4$ and therefore it has a higher
  concentration of mass at its core than the other bodies.
   It is well-known that the internal angular velocity of the Sun varies
  with the radius  and  this requires a modification of the  Clairaut's theory
  \cite{zharkov}. Nevertheless, this radial variation of angular
  velocity seems to be concentrated  near the core
  (see \cite{bi2011} Figure 1) and we decided to apply 
  the usual Clairaut's theory with an averaged angular velocity $\Om$
  in the sense of Tisserand,
  which is  defined by $L=\Io\Om$ where $L$ is the Sun angular
  momentum.
  We  found several proposals of internal density distributions for the Sun
  (Solar Standard Models) and  we did  computations with two of them.

\item[j)] {\it The Sun II.}
  The first density distribution we used is that  in  \cite{burvsa}.
  In this reference the authors provided all constants we needed except
  for $J_{2v}$. Their values are:
  $m=1.9889\,\times\, 10^{30}$kg, $R=696000$km, $\Io=7.60\times 10^{46}$kg m$^2$
  ($\Io/mR^2=0.0785$ and $R_\Irm/R=0.444$), 
  $\Om=2.87\times 10^{-6}$rad/s. The graph of the density function
  used in  \cite{burvsa} is shown in Figure \ref{dep}.
  We used  $J_{2v}=0.2295\times 10^{-6}$ \cite{inpop17a}.
  For this set of data we obtained:
  $\gamma_\Irm=3.5965\times 10^{-6}$s$^{-2}$, $\gamma_{C}/\gamma_\Irm=0.5326$
  (obtained from the numerical integration of Clairaut's equation),
  and $\gm_{ob}/\gm_\Irm=0.7872$. According to Section
  \ref{polysec} the ratio $R_\Irm/R=0.444$ corresponds to
  a polytrope of index $n=2.948$ and a $\gm_P/\gm_\Irm=0.811$. We observe
  that the values $\gamma_{C}/\gamma_\Irm=0.5326$ and $\gm_{ob}/\gm_\Irm=0.7872$
  are very different, indeed $\gm_{ob}$ is closer to the value  $\gm_P$ of the
  polytrope than to $\gamma_{C}$. Since $\frac{\Om^2R^3}{Gm}=0.021\times 10^{-3}$
  is very small, if this density model would be a good representative for the
  real density of the Sun then Clairaut's theory should have given a better result.
  This model has a  larger value of $\Io/mR^2=0.0785$ than others
  found in the literature for which  $\Io/mR^2\approx0.07$. The data for this model
  are not presented in  Tables \ref{tab1} and \ref{tab2}.
  
\item[k)] {\it The Sun III.}
  Since with the density model in \cite{burvsa} we did not get a reasonable result
  we tried a second one that is given in \cite{bahcall1988}. All the results in
  \cite{bahcall1988} are normalized by the Solar radius that we chose
  as $R=695700$km. Explicit values for $m$ and $\Io$ are not provided in
  \cite{bahcall1988}, we obtained them in the following way. Integrating
  the density distribution given in
  \cite{bahcall1988}, and shown in Figure \ref{dep},  we obtained a value
  for the total mass of $1.985549\times 10^{30}$kg.  In order to calibrate
  the total mass to the standard value $m=1.9885\times 10^{30}$kg we multiplied
  the densities provide in \cite{bahcall1988} by the small factor
  $1.9885/1.9855\approx 1.0015$. With this normalized density we computed
  $\Io=6.877\times 10^{46}$kg m$^2$ that implies
  $\Io/mR^2=0.0715$ and $R_\Irm/R=0.423$. In order to obtain $\Om$ we use
  the results in \cite{bi2011} in the following way. In this reference
there is   a graph of the variation of the angular rotation within the Sun as a function
  of the radius (see \cite{bi2011} Figure 1, the model which takes into account
  magnetic effects). This distribution supposes an average  surface velocity of
  $2.9\times\times 10^{-6}$rad/s, we multiplied it by the factor
  $2.87/2.9\approx 0.99$) to obtain the most accepted value
  $\Om=2.87\times\times 10^{-6}$rad/s average angular velocity at the surface
  \cite{beck2000}.
  This changes the estimate $2.02\times 10^{41}$kg m$^2$/s
  for  the solar total angular momentum $L$  in reference
  \cite{bi2011} to $2.00\times 10^{41}$kg m$^2$/s (see \cite{iorio}
  for several other estimates of $L$). Then we defined
  the  average angular velocity $\Om=L/\Io=2.91\times 10^{-6}$rad/s.
  In order to check the consistency of the models used in
  \cite{bahcall1988} and \cite{bi2011}  we computed the
  total angular momentum using the density distribution in
  \cite{bahcall1988} and the varying angular velocity given
  in Figure 1 of \cite{bi2011}, the result is  $2.03\times 10^{41}$kg m$^2$/s
  which is close to the total angular momentum above.
  Integrating numerically Clairaut's equation we obtained
  $\gamma_{C}/\gamma_\Irm= 0.6271$.
  The value  $\gm_{ob}/\gm_\Irm= 0.6313$
  was computed using $\Om=2.87\times\times 10^{-6}$rad/s as
  above and  $J_{2v}=0.2295\times 10^{-6}$ \cite{inpop17a}. Notice that
  $\gamma_{C}/\gamma_\Irm$ and $\gm_{ob}/\gm_\Irm$ are close.
  The polytrope that corresponds to the ratio $R_\Irm/R=0.423$
  has index  $n=3.060$ and  $\gm_P/\gm_\Irm=0.7863$, which is 25\% larger than
  the observed value $\gm_{ob}/\gm_\Irm$. The density function
  of this polytropic approximation normalized to have the same
  $m=1.9885\times 10^{30}$kg
  is shown in Figure \ref{dep}.

\item[l)] {\it The Sun IV.}  
  There are 
    different estimates of  
    $J_{2v}$ for the Sun \cite{rozelot2011}.
        The quantity $\gm_{ob}/\gm_\Irm$ in Table \ref{tab2} 
    is very sensitive to variations of $J_{2v}$ (and also of $\Om$) while 
    $\gm_\Irm$, $\gm_{C}$, $\gm_{P}$, and $\gm_{DR}$ do not depend neither on
    $J_{2v}$ nor on $\Om$. If we fix the quantities
    $m=1.9885\times 10^{30}$kg, $\Io=6.877\times 10^{46}$kg m$^2$
    ($\Io/mR^2=0.0715$ and $R_\Irm/R=0.423$), and $\Om=2.87\times\times 10^{-6}$rad/s
    as in remark (k) and vary $J_{2v}$ from 
    $1.65\times 10^{-7}$ to $7.43\times 10^{-7}$, which are the values 
    in the last three lines of Table 1 of the historical survey \cite{rozelot2011},
    then we obtain  $0.1950<\gm_{ob}/\gm_\Irm<0.8781$ for the variation of
    $\gm_{ob}/\gm_\Irm$. The  lowest value $0.195$
    is close to the lower bound of Theorem \ref{thlow}, which is
    $0.169$.
    If we restrict the variation of $J_{2v}$ to the values from INPOP2008
    $J_2=0.182\times 10^{-6}$ \cite{inpop08} to
    INPOP2017 $J_2=0.2295\times 10^{-6}$\cite{inpop17a}, which was the value
    adopted in this paper, then we obtain $0.6313\le\gm_{ob}/\gm_\Irm<0.7961$.
    Notice that the value $\gm_{ob}/\gm_\Irm=0.7961$, for $J_2=0.182\times 10^{-6}$,
    is close to the value
    $\gm_P/\gm_\Irm=0.7863$ for the polytrope with the same  $R_\Irm/R=0.423$.
     The sensitivity of $\gm_{ob}$
    to  variations of $J_2$, and other parameters as $\Om$ and $\Io$,
    and the empirical difficulty in obtaining a sharp
    estimate of this value explains the variation in our previous 
    determinations of $\gm$ in \cite{rr2015} and \cite{rr2018} and also shows that
    in the future we may be enforced to change our estimate of  $\gm_{ob}$ for
    the Sun again. So, the simple estimate obtained with the
    polytropic approximation
    that  does not match by 25\% the value $\gm_{ob}/\gm_\Irm= 0.6313$,
    which we believe is the best at the moment,
        seems not bad.

\end{itemize}

\begin{table}[hptb]
\begin{tabular}{lllllll}
  \hline\noalign{\smallskip}
  Body & $m\,${\tiny($\times\, 10^{24}\,$kg)}  & $R\,${\tiny(km)} &
  $\frac{\Io}{mR^2}$&$R_\Irm/R$ &
  $\Om$\, {\tiny($\times \, 10^{-5}$s$^{-1})$}&$J_2\,${\tiny($\times 10^{-6})$}\\
  \noalign{\smallskip}\hline\noalign{\smallskip}
Sun$^{(a)}$ &   1988500      &   695700      &0.0715&  0.423     &  0.291
&  0.2295\\
Earth\cite{prem} &   5.974      &   6371      & 0.331& 0.909
&   7.2921      &   1082.6      \\
  Mars\cite{zharkov2005} &  0.6419      &   3390      & 0.365& 0.955      &   7.0882      &   1957.0 \\
  Jupiter\cite{hubbard2016} &   1899      &   69911      &0.264&
0.816      &   17.585      &   14696      \\
  Saturn \cite{nettel2013saturn}
       &   568.3      &   58232  &  0.228  &  0.754    &   16.53 &   16290  \\
Uranus\cite{nettel2013uran} &   86.81      &   25388     &0.227&
0.754      &   10.121      &   3510.7      \\
Neptune\cite{nettel2013uran} &   102.4      &   24622     & 0.238&
0.772      &   10.833      &   3533.0     \\
\noalign{\smallskip}\hline\noalign{\smallskip}
\end{tabular}
\caption{
    $m=$mass, $R=$volumetric mean radius, $\Io$=mean moment of inertia ($[A+B+C]/3$),
    $R_\Irm=$inertial radius defined by $\Io=0.4 mR_\Irm^2$ and
  related to $\Io/(mR^2)$ by
  $\left(\frac{R_\Irm}{R}\right)^2=\frac{5}{2}\frac{\Io}{mR^2}$,
    $\Om$=spin angular velocity,
    $J_2=(C-A)/(mR_e^2)$=dynamic form factor,
    where $C$ is the polar moment of inertia and $A$ is the
    equatorial moment of inertia of the rotating body.
    $^{(a)}$  The constants for the Sun were obtained according to remark (k)
    in the text. The value of $\Om$ for Saturn  is from \cite{mankovich2019cassini}.   
    }
\label{tab1}
\end{table}

\begin{table}[hptb]
\begin{tabular}{llllllc}
  \hline\noalign{\smallskip}
Body& $\gamma_\Irm$ {\tiny ($\times 10^{-6}s^{-2}$)}&
$\gamma_{DR}/\gamma_\Irm$  & $\gamma_{P}/\gamma_\Irm$ &
$\gm_{C}/\gm_\Irm$ & $\gm_{ob}/\gm_\Irm$
&$\frac{\Om^2R^3}{Gm}${\tiny ($\times 10^{-3}$)} \\
  \noalign{\smallskip}\hline\noalign{\smallskip}
  Sun\cite{bahcall1988}&  4.1761  &  -------- &0.7863 & 0.6271 & 0.6314
& 0.021\\
  Earth\cite{prem} &   1.640           &  0.9997      & 1.000 &0.9998 
                   &0.9885 
&  3.449 \\  
Mars\cite{zharkov2005}&
1.009      &  0.9999    &0.9999  & 1.0012 &0.9258
  &4.569      \\
Jupiter\cite{hubbard2016}&
0.5466      &  1.011   & 0.9963  &  0.9954        &0.9802 
&83.39      \\
  Saturn\cite{nettel2013saturn} &  0.3582   &   1.055      & 0.9877 & 0.9650 &

                            0.9924  &   139.0      \\  
  Ura\cite{nettel2013uran}(U1)&
0.6614  &   1.055    & 0.9877 &  -------- &0.9896 
&28.93      \\
Nep\cite{nettel2013uran}(N1)&
0.7974  &     1.037   & 0.9908  & 0.9868     & 0.9805  
&25.63    \\
\noalign{\smallskip}\hline\noalign{\smallskip}
\end{tabular}
\caption{
  $\gm_\Irm=(4/5)(Gm/R_\Irm^3)$,
  $\gamma_{DR}=$value of $\gm$ obtained from  the Darwin-Radau approximation
  (remark (d)),
  $\gm_{P}=$value of $\gm$ for a body  made of a polytropic
  fluid with an index determined by the ratio $R_\Irm/R$ (remark (e)),
  $\gamma_{C}$ is the value of $\gm$ obtained from the numerical
  integration of Clairaut's equation (the density
  functions are shown in Figure \ref{dep}), 
    $\gm_{ob}=\frac{\Io}{mR_e^2}\frac{\Om^2}{J_2}$ (see remark (f)),
$\Om^2R^3/(Gm)$=the centrifugal acceleration at the
    equator over the average gravitational acceleration on its surface
    (small quantity in Clairaut's theory).
  In reference \cite{nettel2013uran}
    there are two density models for Uranus and Neptune, (U1) and (N1) indicate the
    model we used (only the data file for the density of Neptune is provided
    in the supplementary data of \cite{nettel2013uran}, so the value of
    $\gm_C/\gm_{ob}$ was computed only for Neptune).
    For Saturn the difference between $\gm_{C}/\gm_\Irm$ and
    $\gm_{ob}/\gm_\Irm$ may be attributed to corrections of higher order
    in the parameter $\Om^2R^3/(Gm)=0.139$, which in this case
    is  not so small.
    }
\label{tab2}

\end{table}

\section{Conclusion}
\label{conc}

Theorems
\ref{thlow} and \ref{thup}   establish
sharp inequalities for the gravitational  modulus $\gm$ as
a function of the ratio $R_\Irm/R$.
These inequalities, which can be 
useful in the determination of physical properties of exoplanets,
may be improved, or from a practical perspective substituted, 
in the following way.

The  upper bound, $\gm/\gm_\Irm\ge 1.001401$,  in Theorem \ref{thup}  
does not depend on $R_\Irm/R$ because the  geometric radius $R$
can be artificially large due to the presence of  a thick layer of negligible mass.
While this is not a  drawback
for bodies with   $R_\Irm/R$ close to one, which is the case 
for most  planets in the solar system, 
it would be useful to have a more realistic upper (and also a lower) bound
for $\gm/\gm_\Irm$ when $R_\Irm/R<0.7$. One way to improve the inequalities
in Theorems \ref{thlow} and \ref{thlow} would be to  impose additional restrictions
on the density function $\rho$, besides  it being  non increasing, or
on the definition of $R$. Then the analysis of the solutions to
the differential equations (\ref{v2}) and  (\ref{Gm1})
under the new constraints would  give
the desired inequalities. As far as I know  there is  no well-accepted
suggestion of further  restrictions on $\rho$ or on the definition of $R$
in the physical  literature
(see \cite{baschek1991parameters} for a discussion in this direction). 
A way to avoid these general restrictions
is to assume an archetypal  model that we choose as the polytropes.

It has been a practice among researchers,  as for instance Chandrasekhar,
to use polytropes as a first approximation to more realistic stellar models.
As discussed in Section \ref{polysec}, a polytrope is characterized by the polytropic
index $n$ and  by two more parameters that can be the mass and the radius or the
density at the center and the constant $K$. It is remarkable that  $n$ and $R_\Irm/R$
are in one-to-one correspondence and, as shown in Figure \ref{figpolr},
$n\approx \left(1-\frac{R_\Irm}{R}\right) 5$. The gravitational
modulus of a polytrope, denoted as $\gm_P$,   depends only on the index $n$ and
therefore it is determined by the value of $R_\Irm/R$.
For all planets  in Figure \ref{ampl},  except for  Saturn and Uranus,
the value of $\gm_P/\gm_\Irm$ is a good  approximation for  $\gm_C/\gm_\Irm$,
where $\gm_C$ is  the value of $\gm$ obtained
from the integration of Clairaut's equation.
For Saturn the relative difference $|\gm_P-\gm_C|/\gm_P$ is $2.8\%$ and for Uranus
$\gm_C$ was not computed. 
Figure \ref{ampl} also shows that $\gm_P/\gm_\Irm$ is a
very good approximation even for the observed values $\gm_{ob}/\gm_\Irm$
\footnote{For the planets listed in
Tables \ref{tab1} and \ref{tab2}
the largest error of ($|\gm_P/\gm_{ob}-1|$) 
is for Mars, 8\%,  because Mars
may not be in hydrostatic equilibrium. For
the remaining planets  the error is within 1\%.}
with  the small deviations being
possibly explained by the existence of non-gravitational stresses, transient
behavior, or higher order corrections in the small parameter $\Om^2R^3/(Gm)$,
as argued in the second paragraph of Section \ref{computation}.
For the Sun, if $J_2$ is chosen as $0.2295\times 10^{-6}$\cite{inpop17a}, then
$|\gm_P/\gm_{ob}-1|=25\%$; and if $J_2$ is chosen as $0.182\times 10^{-6}$
\cite{inpop08}, then  $|\gm_P/\gm_{ob}-1|=1\%$ (see the vertical line in Figure
\ref{sum} and the remark (l) in Section \ref{computation}).
So within the range
of different values of $J_2$ in the recent literature \cite{rozelot2011}
the value of $\gm_P/\gm_\Irm$ is acceptable even for the Sun. These considerations
lead me to the following:
\begin{itemize}
\item[] {\large \it Practical rule for the estimation of $\gm$}:
  {\it The mass $m$ and the moment of inertia $\Io$ of a large celestial
    body
    determine its
  inertial radius $R_\Irm=\sqrt{\frac{5\Io}{2m}}$ and its square  inertial
  frequency   $\gm_\Irm=(4/5)(Gm/R_\Irm^3)$.
   If in addition
  the volumetric radius $R$ of the body is given, then the ratio $R_\Irm/R$
  and the graph in Figure \ref{figpolr}
  determine the value of $\gm_P/\gm_{\Irm}$ for  a polytrope.
  The gravitational modulus $\gm$ of the body is approximately given by
  $\gm/\gm_\Irm\approx \gm_P/\gm_{\Irm}$. If $R_\Irm/R>0.7$, what happens for the planets in the solar system, then  $\gm/\gm_\Irm\approx \gm_P/\gm_{\Irm}\approx 1$.}
\end{itemize}

\appendix

\section{Appendix: Proofs of some Propositions}

\label{app0}

The following simple result is widely stated in the literature with no proof
or reference.

\begin{proposition}
  \label{RRIin}
  For any spherically symmetric integrable mass density distribution
  $\rho$ with support in $[0,R]$:
  \[
  \left(\frac{R_\Irm}{R}\right)^2\le \frac{5}{3},
  \]
  the value $\left(\frac{R_\Irm}{R}\right)=\frac{5}{3}$ being achieved when
  all the mass is uniformly distributed over a spherical shell of radius $R$.
  If in addition  $\rho$ is non-increasing then
  \[
  \frac{R_\Irm}{R}\le 1.
  \]
In this case,  $R_\Irm/R= 1$ if and only if $\rho$ is constant.  
\end{proposition}
\bproof
The definition of $R_\Irm$  implies 
 \[
   \left(\frac{R_\Irm}{R}\right)^2=
   \frac{5}{3}\frac{\int_0^R a^{2}\frac{a^2}{R^2} \rho(a)  da}{\int_0^R a^2\rho(a)da}
   \le \frac{5}{3}
\]
If all the mass is concentrated on a spherical shell of radius $R$,
$\rho(r)=(m/4\pi R^2)\dt(r-R)$,  then integration gives
$(R_\Irm/R)^2=5/3$.

Now, suppose $\rho$ is non-increasing and let
\[
\hat \rho=\frac{3}{R^3}\int_0^R\rho(r)r^2dr\quad\text{ and}\quad
\rho(r)=\hat\rho +f\Longrightarrow \int_0^Rf r^2dr=0
\]
The signed density  $f$ is not null  if,  and only if,
$\rho$ is not  constant. If $f$ is not null then   there
exists a value $\ov r\in (0,R)$ such that $f(r)\ge 0$ for $r<\ov r$ and
$f(r)\le 0$ for $r>\ov r$ with $\int_0^{\ov r}fr^2dr>0$ and
$\int_{\ov r}^R fr^2dr<0$. These considerations imply that if $f$ is not null:
\[
\begin{split}
  \frac{3}{5}R^2_\Irm =&\frac{\int_0^R a^{4} \rho(a)  da}{\int_0^R a^2\rho(a)da}
  =\frac{\int_0^R a^{4} \hat \rho  da+\int_0^R a^{4} f(a)  da}{\int_0^R a^2\hat \rho da}
\\
=& \frac{3R^2}{5}+\frac{\int_0^{\ov r} a^{4}  f(a)  da+\int_{\ov r}^R a^{4} f(a)  da}
  {\hat \rho R^3/3}\\
  <& \frac{3R^2}{5}+
  \frac{\ov r^2\int_0^{\ov r} a^2  f(a)  da+\ov r^2\int_{\ov r}^R a^2 f(a)  da}
{\hat \rho R^3/3}=\frac{3R^2}{5}
\end{split}
\]
so $(R_\Irm/R)^2<1$. For a body with constant density $\frac{R_\Irm}{R}=1$.
\eproof

We recall the statement of Proposition \ref{stat1}.
  \begin{proposition} Suppose that $\rho$ satisfies hypothesis {\rm (\ref{innerrho2})}.
    Then, for $\Om>0$  there exists a unique bounded solution to equation
    {\rm  (\ref{cl1})} {\rm (}and therefore to problems {\rm  (\ref{cl6})} and
    {\rm  (\ref{cl8})}{\rm )}. This solution is strictly positive, non-decreasing,
    and     $C^1$.
     For $\Om=0$
    the only solution to equation {\rm  (\ref{cl1})}
     is $\ep(r)=0$,  $r\ge 0$.
   \end{proposition}


   In the case $\rho$ is $C^2$ the proof of this result, and more, can be found in
   \cite{poincare1902figures} chapter IV.
   
  In order to solve equation  (\ref{cl1}) we will solve the boundary value
  problem in equations (\ref{cl8}), (\ref{bc8}), and (\ref{jumpcl2}).
  At first we show that any solution to equations
  (\ref{cl8}) with the jump conditions  (\ref{jumpcl2}) imply that $\ep$ is $C^1$.
  Within the intervals $(r_j,r_{j+1})$, $\ep(r)=-\frac{3}{\ov\rho(r)}y$ implies
  \[
  \ep^\prime=-\frac{3}{\ov\rho^2}(y^\prime\ov\rho-y\ov\rho^\prime)=
\frac{3}{r\ov\rho^2}\big(5(y-w)\ov\rho+3y(\rho-\ov\rho)\big),
\]
where we used equations (\ref{cl8}) and (\ref{ovrho2}).
Following the notation in equation (\ref{jumpcl})
 \[
  \Delta \ep^\prime(r_j)=
  \frac{3}{r_j\ov\rho^2(r_j)}\Big(-5[\Delta w(r_j)])\ov\rho(r_j)+3y(r_j)
       [\Delta \rho(r_j)]\Big)=0,
       \]
       where we used $\Delta w(r_j)=\frac{3}{5}\frac{\chi_j}{\ov\rho(r_j)}y(r_j)$
       from  equation (\ref{jumpcl2})
       and $\Delta \rho(r_j)=\chi_j$  from equation (\ref{rhod}). This shows that $\ep$
       is $C^1$.

       Inside the intervals $[r_j,r_{j+1})$ the solution to equation
         (\ref{cl8}) also satisfies equation (\ref{cl5}), namely
\[
r\,\ep^{\, \prime\prime} +  6\,\epsilon^{\,\prime}+
2\,\frac{\ov\rho^\prime}{\ov{\rho}}
\left(r\,\epsilon^{\,\prime}+\epsilon\right)=
(r^6\,\ep^{\, \prime})^\prime+
2\,\frac{r^5\ov\rho^\prime}{\ov{\rho}}
  \left(r\,\epsilon^{\,\prime}+\epsilon\right)=0.
 \]
 At first consider
 the interval $[0,r_1)$ (if $\rho$ is $C^2$ everywhere, then $r_1=\infty$). 
The regularity of $\rho$ at $r=0$ 
 implies that $\rho(r)=\rho(0)+r^2\rho^{\prime\prime}(0)/2+\ldots$ and
 $\ov \rho(r)=\rho(0)+r^2\rho^{\prime\prime}(0) 3/10+\ldots$. So, 
near the origin
equation (\ref{cl5}) can be written as
 \begin{equation}
 \ep^{\prime\prime}+6\frac{\ep^\prime}{r}+
 \bigg(\frac{6\rho^{\prime\prime}(0)}{5\rho(0)}+\Oc(r)\bigg)
 \left(r\,\epsilon^{\,\prime}+\epsilon\right)= 0.
 \label{cl5.1}
 \end{equation}
 If we impose that $\ep$ is bounded
 (twice continuously differentiable)
 at $r=0$,
 then taking the limit as $r\to 0$ into this equation we obtain that 
 $\ep^\prime(0)=0$,  which  implies that near the origin
\begin{equation} 
  \ep(r)=\ep(0)+r^2\ep^{\prime\prime}(0)/2+\ldots\quad\text{where}\quad
  \ep^{\prime\prime}(0)=-\frac{6}{35}\frac{\rho^{\prime\prime}(0)}{\rho(0)}\ep(0).
  \label{bc00}
\end{equation}
In the following we assume that $\ep(0)\ne 0$.
Since $\rho^{\prime\prime}(0)\le 0$, for $r>0$ sufficiently small
$\ep(r)\ep^\prime(r)\ge 0$. If $\ep(r)\ep^\prime(r)\ge 0$ near $r=0$, then
let $\tilde a=\sup_{0<r<r_1}\{\ov\rho^\prime(r)=0\}$. If
$\tilde a=r_1$, then $\ep^\prime(r)= 0$ for $0\le r\le r_1$.
If 
$\tilde a<r_1$, then
$\ov \rho^\prime<0$ in some  interval $(\tilde a,\tilde a +\dt)\subset (0,r_1)$
and  equation (\ref{cl5}) implies that $\ep(r)\cdot\ep^\prime (r)>0$
  in a possibly smaller interval. Now, let
  $\ov a=\sup_{\tilde a<r<r_1}\{\ep^\prime(a)\ne 0\}$ and suppose that $\ov a<r_1$.
   Then equation (\ref{cl5}) implies
  $\ov a\, \ep^{\, \prime\prime}(\ov a)=
   -2\frac{\ov\rho^\prime(\ov a)}{\ov{\rho}(\ov a)} \epsilon(\ov a)$
   and, since $\rho^\prime(\ov a)<0$ and $\ep^2(\ov a)>0$, we obtain
   $\ep(\ov a)\ep^{\, \prime\prime}(\ov a)>0$. But this is impossible because the function
   $F(a)=\ep(a)\ep^\prime(a)$ would be positive for $\tilde a<a<\ov a$ and would
   satisfy
   $F(\ov a)=\ep(\ov a)\ep^\prime(\ov a)=0$ and
   $F^\prime(\ov a)=\ep(\ov a)\ep^{\prime\prime}(\ov a)>0$. So, $\ov a=r_1$ and 
   $\ep(r)\ep^\prime(r)\ge 0$ for $0\le r\le r_1$.
   If $\ep(0)>0$ ($\ep(0)<0$), then $\ep^\prime(r)\ge 0$ ($\ep^\prime(r)\le 0$)
   for $0\le r\le r_1$ and $\ep(r_1)\ge \ep(0)$ ($\ep(r_1)\ge \ep(0)$).
   If $r_1\le R$ is a point
   of discontinuity of $\rho$, then the same argument applied to the interval
   $[0,r_1)$ can be used in the interval $[r_1,r_2)$ to show that 
       $\ep(r)\ep^\prime(r)\ge 0$ for $r_1\le r\le r_2$. The argument can be repeated
       up to the interval $[r_n,\infty)$ to conclude that for any $w_0\ne 0$ and
\begin{equation}         
w(0)=y(0)=w_0\ne 0,\quad \ep(0)=-\frac{3}{\ov\rho(0)}w_0,
\label{iap}
\end{equation}
equation (\ref{cl8}) with the jump conditions in equation (\ref{jumpcl2})
has a solution such that $\ep(r)\ne 0$ and
$\ep(r)\ep^\prime(r)\ge 0$ for $0\le r< \infty$.
In the following we show that the $w$-component of this solution is always different
from zero.

The second and third equations in (\ref{cl8}) imply
\[
(r^5y)^\prime=5r^4w, \quad r^5\frac{\ov \rho \ep}{3}=-r^5y, \quad \text{and}\quad
\frac{d}{dr}\left(r^5\frac{\ov \rho \ep}{3}\right)=-5r^4 w.
\]
This last equation and $\ov\rho^\prime(r)=\frac{3}{r}\bigl[\rho(r)-\ov\rho(r)\bigr]$,
equation (\ref{ovrho2}), imply
\begin{equation}
\frac{2}{15}\ov\rho(r)\ep(r)+\frac{1}{5}\rho(r)\ep(r)+
\frac{r}{15}\ov\rho(r)\ep^\prime(r)=-w(r).
\label{wpos}
\end{equation}
Since  $\ep(r)\ne 0$ and
$\ep(r)\ep^\prime(r)\ge 0$ for $0\le r< \infty$, the 
left hand side of this equation is either strictly   positive 
or  strictly negative.

The boundary value
problem in equations (\ref{cl8}), (\ref{bc8}), and (\ref{jumpcl2})
can be solved with the following  algorithm.
 Let $(\tilde w, \tilde y)$ be the solution to the differential equation
 (\ref{cl8}) with the initial condition $\tilde w(0)=\tilde y(0)=1$ and
 the jump conditions (\ref{jumpcl2}). Since  $\tilde w(R)\ne 0$, 
  the desired solution
  $(w, y)$ to the boundary value problem  is the solution 
  to  equation (\ref{cl8}) with the jump conditions  (\ref{jumpcl2}) that
  satisfies the initial condition
 \[
 w(0)=y(0)=-\frac{\Om^2}{8\pi G}\frac{1}{\tilde w(R)}
 \]

 Since $w(R)\ne 0$ if $w_0\ne 0$, for  $\Om=0$
 the only solution to equation (\ref{cl8})
 with the jump conditions  (\ref{jumpcl2}) that  satisfies the boundary
 condition  $w(R)=0$ in equation (\ref{bc8}) is the trivial solution. This implies that
 the solution to the boundary value problem is unique
 (the difference between two different solutions
 would be a nontrivial solution to the problem with $\Om=0$).

 \section{Appendix: The gravitational potential, 
   the gravitational energy, and the centrifugal energy.}

 \label{app1}

 The main goal in this appendix is to show that the gravitational energy 
of a deformed body
  \begin{equation}
  U(\ep)=\frac{1}{2}
  \int_{\R^3}\tilde\rho(x)\Phi(x)d x, \label{grav2}
\end{equation}
 satisfies
\begin{equation}
  U(\ep)=\frac{32\pi^2 G}{45} V(\ep),
  \label{UV}
\end{equation}
where: $V$ is the functional defined in equation (\ref{V1}),  
$\ep$ is an arbitrary small deformation as those in the theory of Clairaut,
and the equality holds up  to  terms
of order $\ep^2$. We assume that   $\rho$ and $\ep$ are 
$C^2$. Approximation arguments, as those in section \ref{discmass},
may be used to prove  equation (\ref{UV})
under the same hypotheses of Theorem \ref{theor1}.

The interesting identity (\ref{UV}) was pointed out by one of the referees of the paper
and it is non trivial since it requires computations up to the order of $\ep^2$,
instead of only those of order $\ep$ used to  derive Clairaut's equation.
The notation and the theory used in this appendix are those in \cite{zharkov}.

Let $s$ be  the mean radius of the density contour $\tilde \rho(r,\theta)=\rho(s)$,
where $\rho$ is the radial density function of the body  at rest and
$\tilde\rho$ is   density function of a  deformed state. For a given $\theta$
the equation $\tilde \rho(r,\theta)=\rho(s)$ can be solved for $r(\theta,s)$.
If $r(\theta,s)$ is expanded into a sum  of Legendre polynomials $P_l(\cos\theta)$
and only the terms corresponding to $l=0$ and $l=2$ are retained, then 
 (\cite{zharkov} equation 27.8)
\begin{equation}
r(\theta,s)=s\big[1+s_0(s)+s_2(s)P_2(\cos\theta)\big].  
\label{zha27.8}
\end{equation}
Due to  the definition of $s$,  $r(\theta,s)$ must satisfy
 (\cite{zharkov} equation 27.6)
\begin{equation}
  \frac{4\pi}{3}s^3=\int_0^\pi\int_0^{r(\theta,s)}
  \int_0^{2\pi}a^2\sin\theta dad\theta d\phi=\frac{2\pi}{3}
  \int_0^\pi r^3(\theta,s)\sin\theta d\theta.
\label{zha27.6}  
\end{equation}
The identity $\tilde \rho(r,\theta)=\rho(s)$
implies that  equation (\ref{zha27.6}) is  
a necessary condition for the conservation of mass under a  deformation.

In Clairaut's theory, for each given $s\in (0,R]$, the
curve $\theta\to r(\theta,s)$ is an ellipsoid
with small flattening $\ep(s)$, see equation (\ref{ell}).
This implies that up to first order in $\ep(s)$:
$s_0(s)=0$ and $s_2(s)=-\frac{2}{3}\ep(s)$.
For this choice of $s_0$ and $s_2$ equation (\ref{zha27.6}) is verified
up to order $\ep$ but not at order $\ep^2$. So, under deformation,
mass is conserved 
up to order $\ep$ but not at order $\ep^2$. While this is not a problem for
obtaining Clairaut's equation,  which is correct up to order $\ep$, it is a
problem in obtaining the correct expression for the gravitational energy,
which is a quantity of order $\ep^2$. This difficulty is overcome with
the choice $s_0=-\frac{4}{45}\ep^2$, which ensures that equation 
(\ref{zha27.6}) is verified up to order $\ep^2$ but does not change the flattening
of the deformed ellipsoids up to order $\ep$. Therefore equation (\ref{zha27.8})
becomes
\begin{equation}
r(\theta,s)=s\left[1-\frac{2}{3}\ep(s)P_2(\cos\theta)-\frac{4}{45}\ep^2(s)\right].  
\label{rs}
\end{equation}
This equation can be inverted up  to order $\ep^2$ to give
\begin{equation}
  s(\theta,r)=r\left[1+\frac{2}{3}\ep(r)P_2(\cos\theta)+\frac{4}{45}\ep^2(r)
\right]+\frac{2}{9}\frac{d}{dr}[r^2\ep^2(r)] \ [P_2(\cos\theta)]^2.  
\label{rs2}
\end{equation}
This equation and $\tilde \rho(r,\theta)=\rho(s)$ implies that, up to order $\ep^2$,
the density of the deformed body is given by
\begin{equation}
  \begin{split}
    \tilde \rho(r,\theta)=&\rho(r)+\rho^\prime (r)
  \left\{\frac{2}{3}r\ep(r)P_2(\cos\theta)+\frac{4}{45}r\ep^2(r)+
    \frac{2}{9}\frac{d}{dr}[r^2\ep^2(r)] \ [P_2(\cos\theta)]^2\right\}\\
  &+\frac{1}{2}\rho^{\prime\prime} (r)\frac{4}{9} r^2\ep^2(r) [P_2(\cos\theta)]^2.
\end{split}
\label{rhot}
\end{equation}

In order to compute the  gravitational potential of the deformed body
\[
  \Phi(x)=-G\int_{\R^3}\frac{\tilde \rho(\tilde x)}{|x-\tilde x|} d\tilde x
\]
 we use
the expansions
\[
  \int_0^{2\pi}\frac{d\phi}{|x-\tilde x|}=
  \frac{2\pi}{r}\sum_{n=0}^\infty \left(\frac{\tilde r}{r}\right)^n
  P_n(\cos\theta)P_n(\cos\tilde\theta)\quad
\text{for}\quad r>\tilde r 
\]
and
\[
\int_0^{2\pi}\frac{d\phi}{|x-\tilde x|}=\frac{2\pi}{\tilde r}
\sum_{n=0}^\infty \left(\frac{r}{\tilde r}\right)^nP_n(\cos\theta)P_n(\cos\tilde\theta)\quad
\text{for}\quad \tilde r > r. 
\]
Then, from the orthogonality relations of the Legendre polynomials and from
equation (\ref{rhot}),  we obtain
\begin{equation}
  \begin{split}
    & \Phi(r,\theta)=\Phi_0(r)+\tilde\Phi_0(r)+
    [\Phi_2(r)+\Oc(\ep^2)] P_2(\cos\theta)+\Oc(\ep^3)\qquad\text{where}\\
   & \Phi_0(r)=  -4\pi G\left\{\frac{1}{r}\int_{0}^r a^2 \rho(a)da
     +\int_{r}^\infty a \rho(a)da\right\}\\
   & \tilde \Phi_0(r)=-\frac{8\pi}{45} G\int_{r}^\infty a^2 \rho^\prime(a)
   \ep^2(a)da \\
  & \Phi_2(r)=  -\frac{8\pi}{15} G\left\{\frac{1}{r^3}
  \int_{0}^r a^5\ep(a) \rho^\prime(a)da
  +r^2\int_{r}^\infty \ep(a) \rho^\prime(a)da\right\}
\end{split}
\label{phi12}
\end{equation}
where we used $\int_0^\pi [P_2(\cos\theta)]^3 \sin\theta d\theta=4/35$.
Notice that $\Phi_0=\Oc(\ep^0)$ and $\Phi_2=\Oc(\ep)$ are the functions
in equation (\ref{phi1}) that are used in the derivation of Clairaut's equation.

The gravitational energy in equation (\ref{grav2})  can be written as
 \begin{equation}
   U=
  \pi\int_0^\pi\int_0^\infty \tilde\rho(r,\theta)\left[
\Phi_0(r)+\tilde\Phi_0(r)+
    [\Phi_2(r)+\Oc(\ep^2)] P_2(\cos\theta)\right] \sin\theta d\theta \frac {d r^3}{3}
\label{U2}
\end{equation}
where terms of order $\Oc(\ep^3)$ were neglected. In order to compute this
integral it is convenient to perform the change of variables $r=r(\theta,s)$ in
equation (\ref{rs})  that implies
\[
  \begin{split}
    \tilde\rho(r,\theta)&= \rho(s)\\
    d r^3\, d\theta &=\Big\{ ds^3-2P_2(\cos\theta)d[s^3\ep(s)]
    -\frac{4}{15}d[s^3\ep^2(s)]\\
    &\qquad\qquad\qquad +\frac{4}{3}[P_2(\cos\theta)]^2d[s^3\ep^2(s)]
    +\Oc(\ep^3)\Big\}d\theta\\
    \Phi_0(r)&=\Phi_0(s)-\Phi_0^\prime(s)\frac{2}{3}P_2(\cos\theta)s\ep(s)\\
   &\qquad +\left\{ \Phi_0^{\prime\prime}(s)\frac{2}{9}[P_2(\cos\theta)]^2s^2
     - \Phi_0^\prime(s)\frac{4}{45} s\right\}\ep^2(s) +\Oc(\ep^3)\\
      \tilde \Phi_0(r)&= \tilde \Phi_0(s)+\Oc(\ep^3)\\
      \Phi_2(r)&= \Phi_2(s)-\Phi^\prime_2(s)\frac{2}{3}P_2(\cos\theta)s\ep(s)
      +\Oc(\ep^2)
     \end{split}
 \]
 Using all these relations the integral in equation (\ref{U2}) can be computed
 up to order $\ep^2$ and, after a long computation,  it gives
 \[
   U=
   \frac{4\pi}{15}\int_0^R
   \rho^\prime(s)\ep(s)s^3\left[\Phi_2(s)-
     \Phi^\prime_0(s)\frac{2}{3}s\ep(s)\right]ds
\]
This equation and equation (\ref{V1}) imply equation (\ref{UV}).

The centrifugal potential is given by $\Phi_c=-\frac{\Om^2}{2}(x_1^2+x_2^2)$
and the centrifugal  energy by
 \begin{equation}
  U_c(\ep)=
  \int_{\R^3}\tilde\rho(x)\Phi_c(x)d x=-
  \Om^2\pi\int_0^\pi\int_0^\infty \tilde\rho(r,\theta)\,\,
  r^2\sin^2\theta \,\, r^2\sin\theta dr
\label{Uc}
\end{equation}
Again  it is convenient to perform the change of variables $r=r(\theta,s)$ given in
equation (\ref{rs}). Using that $\sin^2\theta=2[1-P_2(\cos\theta)]/3$ and
 $\Io=\frac{8\pi}{3}\int_0^R\rho a^4da$, a computation
keeping terms  up to order $\ep$ gives 
\[
  U_c(\ep)=-\frac{\Om^2}{2}\Io-\Om^2\frac{8\pi}{45}\int_0^R\rho(s)d[s^5\ep(s)]
\]
The first term, which  does not depend on the deformation, will be neglected.
So
\begin{equation}
  U_c(\ep)=-\Om^2\frac{8\pi}{45}\int_0^R\rho(s)d[s^5\ep(s)]=
  -\Om^2\frac{8\pi}{45}\langle 1,\ep\rangle_\rho
\label{Uc2}
\end{equation}
where the inner product $\langle\cdot,\cdot\rangle_\rho$ is defined in equation
(\ref{innerrho}).

\nd {\it Acknowledgments.}
This paper is dedicated to Prof. Waldyr Muniz Oliva
who first taught me about the theory of figures \cite{Oliva}.
I thank Pedro Tonelli for helping me with the theory of optimal
control which I tried to use to prove Theorem  \ref{thup}.
I specially acknowledge one of the referee's who
gave a great contribution to this paper.

\bibliographystyle{plain}

\begin{thebibliography}{10}

\bibitem{bahcall1988}
John~N Bahcall and Roger~K Ulrich.
\newblock Solar models, neutrino experiments, and helioseismology.
\newblock {\em Reviews of Modern Physics}, 60(2):297, 1988.

\bibitem{baschek1991parameters}
B~Baschek, M~Scholz, and R~Wehrse.
\newblock The parameters $r$ and $t_{eff}$ in stellar models and observations.
\newblock {\em Astronomy and Astrophysics}, 246:374--382, 1991.

\bibitem{beck2000}
John~G Beck.
\newblock A comparison of differential rotation measurements--(invited review).
\newblock {\em Solar physics}, 191(1):47--70, 2000.

\bibitem{bi2011}
SL~Bi, TD~Li, LH~Li, and WM~Yang.
\newblock Solar models with revised abundance.
\newblock {\em The Astrophysical Journal Letters}, 731(2):L42, 2011.

\bibitem{burvsa}
Milan Bur{\v{s}}a, Ladislav K{\v{r}}ivsk{\`y}, and Ond{\v{r}}ejka
  Hovorkov{\'a}.
\newblock Gravitational potential energy of the sun.
\newblock {\em Studia Geophysica et Geodaetica}, 40(1):1--8, 1996.

\bibitem{chandra1963}
S~Chandrasekhar and PH~Roberts.
\newblock The ellipticity of a slowly rotating configuration.
\newblock {\em The Astrophysical Journal}, 138:801, 1963.

\bibitem{chandra1933}
Subrahmanyan Chandrasekhar.
\newblock The equilibrium of distorted polytropes iv.
\newblock {\em MNRAS}, 93:539--574, 1933.

\bibitem{cook2009interiors}
Alan~H Cook.
\newblock Interiors of the planets.
\newblock {\em Interiors of the Planets, by AH Cook, Cambridge, UK: Cambridge
  University Press, 2009}, 2009.

\bibitem{rr2018}
ACM Correia, C~Ragazzo, and LS~Ruiz.
\newblock The effects of deformation inertia (kinetic energy) in the orbital
  and spin evolution of close-in bodies.
\newblock {\em Celestial Mechanics and Dynamical Astronomy}, 130(8):51, 2018.

\bibitem{prem}
Adam~M Dziewonski and Don~L Anderson.
\newblock Preliminary reference earth model.
\newblock {\em Physics of the earth and planetary interiors}, 25(4):297--356,
  1981.

\bibitem{inpop08}
A~Fienga, J~Laskar, T~Morley, H~Manche, P~Kuchynka, C~Le~Poncin-Lafitte,
  F~Budnik, M~Gastineau, and L~Somenzi.
\newblock Inpop08, a 4-d planetary ephemeris: from asteroid and time-scale
  computations to esa mars express and venus express contributions.
\newblock {\em Astronomy \& Astrophysics}, 507(3):1675--1686, 2009.

\bibitem{helled2017internal}
Ravit Helled and Tristan Guillot.
\newblock Internal structure of giant and icy planets: importance of heavy
  elements and mixing.
\newblock {\em Handbook of Exoplanets}, pages 1--19, 2017.

\bibitem{hubbard2016}
WB~Hubbard and B~Militzer.
\newblock A preliminary jupiter model.
\newblock {\em The Astrophysical Journal}, 820(1):80, 2016.

\bibitem{iorio}
Lorenzo Iorio.
\newblock Constraining the angular momentum of the sun with planetary orbital
  motions and general relativity.
\newblock {\em Solar Physics}, 281(2):815--826, 2012.

\bibitem{kellermann2018interior}
Clemens Kellermann, Andreas Becker, and Ronald Redmer.
\newblock Interior structure models and fluid love numbers of exoplanets in the
  super-earth regime.
\newblock {\em Astronomy \& Astrophysics}, 615:A39, 2018.

\bibitem{Lamb}
H.~Lamb.
\newblock {\em Hydrodynamics}.
\newblock Cambridge Mathematical Library, Cambridge, 6th edition, 1932.

\bibitem{mankovich2019cassini}
Christopher Mankovich, Mark~S Marley, Jonathan~J Fortney, and Naor Movshovitz.
\newblock Cassini ring seismology as a probe of saturn’s interior. i. rigid
  rotation.
\newblock {\em The Astrophysical Journal}, 871(1):1, 2019.

\bibitem{mckenzie1966}
Dan~P McKenzie.
\newblock The viscosity of the lower mantle.
\newblock {\em Journal of Geophysical Research}, 71(16):3995--4010, 1966.

\bibitem{moritz}
Helmut Moritz.
\newblock The figure of the earth: theoretical geodesy and the earth's
  interior.
\newblock {\em Karlsruhe: Wichmann, c1990.}, 1990.

\bibitem{nettel2013uran}
N~Nettelmann, R~Helled, JJ~Fortney, and R~Redmer.
\newblock New indication for a dichotomy in the interior structure of uranus
  and neptune from the application of modified shape and rotation data.
\newblock {\em Planetary and Space Science}, 77:143--151, 2013.

\bibitem{nettel2013saturn}
Nadine Nettelmann, Robert P{\"u}stow, and Ronald Redmer.
\newblock Saturn layered structure and homogeneous evolution models with
  different eoss.
\newblock {\em Icarus}, 225(1):548--557, 2013.

\bibitem{Oliva}
W.M. Oliva.
\newblock Massas fluidas em rota\c{c}\~ao e os elipsoides de {R}iemann, uma
  abordagem informal.
\newblock {\em Revista Matem\'atica Universit\'aria}, 47:28--37, 2007.

\bibitem{poincare1902figures}
Henri Poincar{\'e}.
\newblock {\em Figures d'{\'e}quilibre d'une masse fluide: le{\c{c}}ons
  profess{\'e}es {\`a} la Sorbonne en 1900}, volume~13.
\newblock C. Naud, 1902.

\bibitem{rr2015}
C~Ragazzo and LS~Ruiz.
\newblock Dynamics of an isolated, viscoelastic, self-gravitating body.
\newblock {\em Celestial Mechanics and Dynamical Astronomy}, 122(4):303--332,
  2015.

\bibitem{rr2017}
C~Ragazzo and LS~Ruiz.
\newblock Viscoelastic tides: models for use in celestial mechanics.
\newblock {\em Celestial Mechanics and Dynamical Astronomy}, 128(1):19--59,
  2017.

\bibitem{rau1974variational}
ARP Rau.
\newblock Variational principles for the ellipticity of slowly rotating
  configurations.
\newblock {\em Monthly Notices of the Royal Astronomical Society},
  168(2):273--286, 1974.

\bibitem{reed1980}
M~Reed and B~Simon.
\newblock Methods of modern mathematical physics (revised ed.), volume i:
  Functional analysis, 1980.

\bibitem{Rochester-Smylie}
M.~G. Rochester and D.~E. Smylie.
\newblock On changes in the trace of the {E}arth's inertia tensor.
\newblock {\em J. Geophys. Res.}, 79:4948--4951, 1974.

\bibitem{rozelot2011}
J-P Rozelot and C~Damiani.
\newblock History of solar oblateness measurements and interpretation.
\newblock {\em The European Physical Journal H}, 36(3):407--436, 2011.

\bibitem{snellen2014fast}
Ignas~AG Snellen, Bernhard~R Brandl, Remco~J de~Kok, Matteo Brogi, Jayne
  Birkby, and Henriette Schwarz.
\newblock Fast spin of the young extrasolar planet $\beta$ pictoris b.
\newblock {\em Nature}, 509(7498):63, 2014.

\bibitem{stakgold}
Ivar Stakgold.
\newblock {\em Boundary Value Problems of Mathematical Physics: 2-Volume Set},
  volume~29.
\newblock Siam, 2000.

\bibitem{inpop17a}
V~Viswanathan, A~Fienga, M~Gastineau, and J~Laskar.
\newblock Inpop17a planetary ephemerides.
\newblock {\em Notes Scientifiques et Techniques de l’Institut de
  m{\'e}canique c{\'e}leste,(ISSN 1621--3823),\# 108, ISBN 2-910015-79-3, 2017,
  39 pp.}, 108, 2017.

\bibitem{yoder1995astrometric}
Charles~F Yoder.
\newblock Astrometric and geodetic properties of earth and the solar system.
\newblock {\em Global Earth Physics: A Handbook of Physical Constants},
  1:1--31, 1995.

\bibitem{zeidler1}
E~Zeidler.
\newblock Nonlinear functional analysis and its applications i.(fixed point
  theorems) 1986.

\bibitem{zharkov2005}
VN~Zharkov and TV~Gudkova.
\newblock Construction of martian interior model.
\newblock {\em Solar System Research}, 39(5):343--373, 2005.

\bibitem{zharkov}
VN~Zharkov and VP~Trubitsyn.
\newblock Physics of planetary interiors, ed.
\newblock {\em WB Hubbard. Tucson, AZ: Pachart}, 1978.

\end{thebibliography}

\end{document}